\documentclass[aps,prb,showpacs,
twocolumn,floats,superscriptaddress,10pt]{revtex4-1}

\usepackage{dcolumn}
\usepackage[pdftex]{graphicx}
\usepackage[utf8]{inputenc} 
\usepackage[UKenglish]{babel}
\usepackage{amsmath,amssymb,euscript}

\usepackage{ifthen}

\RequirePackage[
   hyperindex,colorlinks,bookmarksnumbered,
   plainpages=true,pdfstartview=FitH]{hyperref}
\hypersetup{linkcolor=blue,urlcolor=blue,citecolor=blue}
\usepackage{hyperref}

\newcommand{\hideandshow}[1]{%
 \ifthenelse{\isundefined{\showme}}{}{#1}}
\newcommand{\showandhide}[1]{%
 \ifthenelse{\isdefined{\showme}}{}{#1}}

\newcommand{\ttimes}{{ \times }}
\newcommand{\dloc}{d_{\rm loc}}

\newcommand{\dbase}{d_{\rm f}} 
\newcommand{\Ntot}{N_{\rm tot}}
\newcommand{\Ntotmult}{N_{\rm tot}^{*}}
\newcommand{\Ntotx}[1]{N_{\rm tot}^{*,#1}}
\newcommand{\Nflavor}{m}
\newcommand{\Nkept}{N_{\rm{K}}}
\newcommand{\Nkeptmult}{N_{\rm{K}}^{*}}
\newcommand{\Nchan}{N_{\rm{c}}} 
\newcommand{\w}{\omega}  
\newcommand{\tildet}{{\tilde t}}  
\newcommand{\NkeptNflavor}{\Nkept^{(\Nflavor)}} 
\newcommand{\Etrunc}{E_{\rm trunc}}
\newcommand{\Eabstrunc}{E_{\rm abs-trunc}}
\newcommand{\geom}{{\rm geom}}
\newcommand{\Fig}[1]{Fig.~\ref{#1}} 
\newcommand{\Figs}[1]{Figs.~\ref{#1}} 
\def\myempty{myemptytoken}
\newcommand{\Eq}[2][\myempty]{%
  \ifx\myempty#1{Eq.~\eqref{#2}}%
  \else{Eq.~(\ref{#2}#1)}%
  \fi
 }
\newcommand{\Sec}[1]{Sec.~\ref{#1}} 
\newcommand{\charge}{{\rm charge}}
\newcommand{\chargechannel}{{\rm charge,channel}} 
\newcommand{\parthole}{{\rm charge}}

\newcommand{\channel}{{\rm channel}}

\newcommand{\spin}{{\rm spin}}
\newcommand{\sNRG}{{\rm sNRG}}
\newcommand{\iNRG}{{\rm iNRG}}

\begin{document}

\title{Interleaved numerical renormalization group
as an efficient multiband impurity solver}

\author{K. M. Stadler} \affiliation{Physics
  Department, Arnold Sommerfeld Center for Theoretical Physics and
  Center for NanoScience, Ludwig-Maximilians-Universit\"at M\"unchen,
  80333 M\"unchen, Germany} 
\author{A. K. Mitchell} \affiliation{Institute for Theoretical Physics, 
  Utrecht University, Leuvenlaan 4, 3584 CE Utrecht, The Netherlands} 
\author{J. von Delft} \affiliation{Physics
  Department, Arnold Sommerfeld Center for Theoretical Physics and
  Center for NanoScience, Ludwig-Maximilians-Universit\"at M\"unchen,
  80333 M\"unchen, Germany}
 \author{A. Weichselbaum} \affiliation{Physics
  Department, Arnold Sommerfeld Center for Theoretical Physics and
  Center for NanoScience, Ludwig-Maximilians-Universit\"at M\"unchen,
  80333 M\"unchen, Germany}  

\date{Received 12 February 2016; revised manuscript received 24 March 2016; published 1 June 2016}
\begin{abstract}
  Quantum impurity problems can be solved using the
  numerical renormalization group (NRG), which involves
  discretizing the free conduction electron system and
  mapping to a `Wilson chain'.  It was shown recently that
  Wilson chains for different electronic species can be
  interleaved by use of a modified discretization,
  dramatically increasing the numerical efficiency of the RG
  scheme [Phys.\ Rev.\ B \underline{89}, 121105(R) (2014)].
  Here we systematically examine the accuracy and efficiency
  of the `interleaved' NRG (iNRG) method in the context of
  the single impurity Anderson model, the two-channel Kondo
  model, and a three-channel Anderson-Hund model. The
  performance of iNRG is explicitly compared with `standard'
  NRG (sNRG): when the average number of states kept per
  iteration is the same in both calculations, the accuracy
  of iNRG is  equivalent to that of sNRG but the
  computational costs are significantly lower in iNRG when
  the same symmetries are exploited. Although iNRG weakly
  breaks SU($N$) channel symmetry (if present), both accuracy
  and numerical cost are  entirely competitive with sNRG
  exploiting full symmetries. iNRG is therefore shown to be
  a viable and technically simple alternative to sNRG for
  high-symmetry models. Moreover, iNRG can be used to solve
  a range of lower-symmetry multiband problems that are
  inaccessible to sNRG.
\end{abstract}

\maketitle


\section{Introduction and Motivation}
\label{sec:intro}

Quantum impurity problems are relevant to a range of
physical phenomena in which strong electron correlations
play a key role.\cite{hewson} They describe a generic class
of systems comprising a few interacting degrees of freedom
coupled to a continuum bath of non-interacting conduction
electrons. The Kondo model\cite{kondo} is the simplest
exemplar, featuring a single spin-$\tfrac{1}{2}$ `impurity'
coupled to a single spinful conduction electron channel. The
basic physics can be understood within the renormalization
group (RG) framework: the effective impurity-bath coupling
grows as the temperature/energy scale is reduced. The RG
flow from weak to strong coupling is characterized by the
Kondo temperature $T_K$, which sets the scale for onset of
strong coupling physics and the dynamical screening of the
impurity spin by conduction electrons.\cite{hewson}

A detailed understanding of this problem was first obtained
using Wilson's numerical renormalization group
(NRG).\cite{wilson,kww1980,nrg:rev} The method involves
discretization of the conduction electron Hamiltonian, and
mapping to a 1D tight-binding `Wilson chain'. The
transformation is defined so that the interacting impurity
subsystem couples to one end of the non-interacting Wilson
chain. A special form of the discretization is used that
ensures exponential decay of hopping matrix elements down
the chain.\cite{wilson} This energy-scale separation
justifies an RG scheme based on successive diagonalization
and truncation, starting at the impurity subsystem and
working down the chain. At each step, a Wilson shell with
$\dloc$ additional local quantum degrees of freedom couples
into the system, but only the lowest $\Nkept$ eigenstates of
the enlarged state space are kept after diagonalization.
This scheme ensures that the Fock space of kept states does
not increase exponentially with chain length, and allows the
physics to be investigated at successively lower energies.

The computational costs of using NRG scale
\emph{exponentially} with the number of fermionic bands
(distinct flavors), $\Nflavor$, involved in the quantum
impurity model. The power and applicability of NRG would be
greatly improved if these numerical costs could be reduced,
since \emph{multi-flavor} quantum impurity problems appear
in a wide range of contexts.  For example, iron impurities
in gold are described by a spin-$\tfrac{3}{2}$ three-channel
Kondo model;\cite{FeinAu,Hanl2013} multiple impurities
separated in real-space\cite{jones,multiimp} or manipulated
by STM\cite{bork} necessitate a multi-channel description,
as do magnetic nanostructures;\cite{magnano} single carbon
nanotube quantum dots display entangled spin-orbital SU(4)
Kondo physics,\cite{su4kondo} while certain nanotube double
dot\cite{cntdqd} and multi-lead semiconductor coupled dot
devices\cite{2ckinqd,akm:tqd2ch,akm:oddimp} are described by
generalized two-channel models; and nanowire/superconductor
heterostructures hosting lead-coupled Majorana fermions give
rise to effective multi-channel topological Kondo
models.\cite{bericooper,akm:MF} Furthermore, quantum
impurity problems appear as effective local models within
dynamical mean-field theory (DMFT) for correlated materials.
Multi-orbital/band lattice models map to generalized
multi-channel impurity
problems,\cite{dmft_rev,asymtrunc,stadler2015dmft} and in
cluster extensions of DMFT, the number of bands of the
effective impurity model scales with the number of cluster
sites.\cite{DCA_dmft} 

There is thus much incentive to improve the efficiency of
NRG when dealing with multi-flavor models. The present
paper aims to make a contribution towards this goal, by
offering a detailed analysis of a recently-proposed scheme
of `interleaving' the Wilson chains for different fermion
flavors.\cite{Mitchell2014} Having a purely methodological
focus, it is based on well-studied physical models and is
particularly addressed at a readership of NRG
practitioners. New physical applications of iNRG are left
for follow-up projects.

To set the scene, we first briefly summarize why the
numerical costs of NRG scale exponentially with $\Nflavor$.
For a given conduction electron discretization, the
\textit{accuracy} of the calculation is controlled by the
number of states \textit{retained} or \textit{kept} at each
step of the iterative RG scheme, $\Nkept$.  On the other
hand, the computational \textit{cost} of an NRG run is
controlled by the \emph{total size} of the Hilbert space to
be diagonalized at each step, $ \Ntot = \Nkept \ttimes \dloc
$, which is the tensor product of the space of states
retained from the previous iteration (of dimension
$\Nkept$), and the state space of a newly added Wilson shell
(of dimension $\dloc$).  The computational time for matrix

diagonalization scales as $\Ntot^3$,  
while the memory required
scales as $\Ntot^2$. 
In Wilson's original `standard' NRG
formulation\cite{wilson} (sNRG), the local dimension for a system with
$\Nflavor$ distinct fermionic flavors scales exponentially  in $m$,
$\dloc^{\rm{\sNRG}}=\dbase^{\Nflavor}$, with $\dbase$ the state 
space dimension of a single flavor.
For a single fermionic level it follows that $\dbase=2$, since it
can be either occupied or unoccupied.  Commonly, quantum
impurity models involve $\Nchan$ channels of \emph{spinful}
conduction electrons. In this case, $\Nflavor=2\Nchan$, such
that $\dloc^{\rm{\sNRG}}=4^{\Nchan}$.

In fact, as the 
number $\Nflavor$ of flavors
increases, the number $\Nkept$ 
of states kept at each step of an NRG calculation must also
be increased to maintain the same accuracy (i.e.,\ the same
degree of numerical convergence).  We find that for
converged sNRG calculations, $\Nkept$ scales roughly
exponentially with the number of flavors, which we will
indicate by writing $\Nkept\equiv \NkeptNflavor$. This
scaling property is demonstrated explicitly in this paper.

Overall then, $\Ntot$ depends exponentially on $\Nflavor$
through both $\Nkept$ and $\dloc$ in sNRG:
\begin{eqnarray}
   \label{eq:Ntot_NRG}
   \Ntot^\sNRG = \NkeptNflavor \ttimes\, \dbase^{\Nflavor} \;.
\end{eqnarray}
This exponential scaling imposes severe limitations on the
applicability of sNRG to treat quantum impurity problems
with several conduction electron channels. In practice,
unless large symmetries can be exploited, sNRG cannot be
used for problems with more than two spinful channels.

Two approaches have been developed to improve the efficiency
of NRG applied to multi-channel quantum impurity models. One
approach exploits non-Abelian symmetries if present:
diagonalization of the NRG Hamiltonian at each step can then
be done in \emph{multiplet} space rather than state space,
significantly reducing the matrix sizes and hence
computational cost.

From the very first sNRG studies of the Anderson impurity
model,\cite{kww1980} it was essential to exploit the SU(2) spin
symmetry so that the calculations could be performed with the
limited computational resources available at that
time.
In Ref.~\onlinecite{Toth2008}, the use of SU(2)
symmetries was incorporated into the framework of the
density-matrix (DM) NRG,\cite{Hofstetter2000} to obtain
dynamical results for a symmetric two-channel model.
Finally, a generalized and flexible framework was pioneered
in Ref.~\onlinecite{Weichselbaum2012}, which now allows much
larger symmetries to be handled, including arbitrary
non-Abelian symmetries.  The precise gain in computational
efficiency with this scheme naturally depends on the
specific model and its symmetries; its scope of application
is of course limited when symmetry-breaking perturbations
(such as a magnetic field) are present.

A second, very different strategy has recently been proposed
in Ref.~\onlinecite{Mitchell2014}.  This `interleaved' NRG
(iNRG) method, described in detail in
Sec.~\ref{sec_Methods}, introduces slightly different
discretization schemes for conduction bands of different
electronic flavors, leading to inequivalent Wilson chains
(even for flavors related by symmetries of the bare model).
For $\Nflavor$ electronic flavors, the $\Nflavor$ Wilson
chains are interleaved to form a single generalized Wilson
chain,\cite{Mitchell2014} which still has the required
property of exponential energy-scale separation down the
chain. The diagonalization and truncation step in iNRG is
then done \textit{separately} after addition of
\textit{each} electron flavor, rather than after addition of
the entire `shell' of $\Nflavor$ flavors, as in \sNRG. In
practice, we specify the truncation threshold not by fixing
the number of states to be kept, but by fixing a truncation
energy: all states with higher energies are discarded at
every step.

Full interleaving
leads to a reduction of the local
state space from $\dloc^{\sNRG}=\dbase^{\Nflavor}$ in sNRG
to $\dloc^{\iNRG}=\dbase$ in iNRG, independent of
$\Nflavor$.  However, it also raises the question as to
whether the truncation energy required to reach accurate,
well-converged results needs to be changed when switching
from sNRG to iNRG.  One of the main conclusions of the
present paper is that it essentially does not change: an
extensive comparison of iNRG and sNRG results, obtained
using comparable discretization settings and exploiting the
same symmetries for both methods, shows that results of
comparable accuracy are obtained if on average the `same'
truncation energy is used (see \Sec{sec:Etrunc-1} for a
detailed discussion).  Moreover, this implies that the
number of states kept at a given step is the same, on
average, for both methods: \begin{eqnarray}
\label{eq:Nkept=Nkept} \Nkept^\iNRG \simeq
\Nkept^\sNRG\equiv\NkeptNflavor \, .  \end{eqnarray} We find
that $\NkeptNflavor$  still depends exponentially on
$\Nflavor$, as for sNRG.  Thus, for iNRG, the computational
costs are governed by \begin{eqnarray} \label{eq:Ntot_iNRG}
\Ntot^\iNRG   = \NkeptNflavor \ttimes \dbase \, ,
\end{eqnarray} where the first factor $\NkeptNflavor$ is
essentially the same as that in \Eq{eq:Ntot_NRG} for
$\Ntot^\sNRG$.  However, the exponential dependence of
$\dloc$ on $\Nflavor$ in the second factor is entirely
eliminated in iNRG.

As a result, when equivalent settings are used for both
methods, iNRG yields results of comparable accuracy as sNRG
at dramatically reduced numerical cost: computation times
are smaller by a factor of order $( \Ntot^\sNRG/\Ntot^\iNRG
)^3=\dbase ^{3(\Nflavor-1)}$, and the required storage
resources are smaller by a factor of order $\dbase
^{2(\Nflavor-1)}$.

Although $\dloc$ is smaller in iNRG than sNRG, an additional
minor cost is incurred in iNRG because the interleaved
Wilson chain is $\Nflavor$ times longer than the standard
Wilson chain (resulting in an additional \textit{linear}
increase in overall computation time with $\Nflavor$).
Furthermore, fine tuning of bare parameters is also
necessary for effective restoration of broken symmetries in
cases where flavor symmetry-breaking is a relevant
perturbation, requiring multiple iNRG runs (the
exponentially rapid convergence in the number of runs is
discussed in Sec.~\ref{sec_tuning}).

The conclusions summarized above are established in this
paper by a direct comparison of iNRG and sNRG for several
symmetric quantum impurity problems (specified in
\Sec{sec_Models}) with $\Nchan=1,2,$ and $3$ spinful
conduction electron channels.  Within iNRG, we explore
different ways of interleaving the electronic flavors, and
exploit all symmetries that remain after interleaving. For
each such iNRG calculation, we perform a corresponding sNRG
calculation using the same symmetries, a comparable
discretization choice, and the same average truncation
energies -- i.e.\ we adopt `equivalent settings'.
Moreover, for each model, we also perform a set of benchmark
calculations exploiting the \textit{full} symmetries of the
bare model, serving as an absolute reference.

Our iNRG-sNRG comparison for equivalent settings focuses
particularly on comparing their \textit{efficiency}
(\Sec{sec_results_efficiency}) and their \textit{accuracy}
(\Sec{sec_results_accuracy}).  We determine efficiency by
tracking representative CPU times.  We gauge accuracy in two
ways: (i) deviations of numerically computed physical
quantities from certain exact results yield an absolute
measure of the accuracy of both methods;  (ii) the discarded
weight\cite{Weichselbaum2011} estimates the degree of
numerical convergence of a given NRG run (see also
\Sec{sec:discweight}).

It may be surprising at first that the accuracy of iNRG and
sNRG are equivalent when using equivalent settings, since
iNRG involves significantly more truncation steps. This
result can, however, be rationalized by noting that the
truncation at each step of iNRG is less severe than in sNRG
(fewer states are discarded at any given step), producing a
more fine-grained RG description.  For equivalent settings,
iNRG clearly outperforms sNRG in terms of efficiency because
the state space diagonalized at each step is much smaller
in iNRG. In fact, for the high-symmetry multiband models
studied here, iNRG is absolutely competitive even when
compared to sNRG calculations that exploit the \textit{full}
symmetry of the model.

This finding greatly increases the scope of possibilities
available for NRG treatments of multiband impurity models.
For models with high symmetries, \emph{both} sNRG and iNRG
can be highly efficient methods.  In such cases, iNRG is a
viable and technically simple alternative to sNRG.  For
models having lower symmetries (for example, when a magnetic
field is applied, particle-hole symmetry is broken, or other
channel anisotropies are present), iNRG has a clear
advantage over sNRG.

In a pure renormalization group (RG) sense, the artificial
  symmetry breaking, of course, is clearly also visible in the
  resulting energy flow diagrams derived from finite-size
  spectra.\cite{wilson,kww1980,nrg:rev} There, a full RG step,
  which in sNRG requires two iterations (e.g.,\ to get from one
    even site to the next even site), now requires $2m$ iNRG steps.
  Nevertheless, aside from possible fine tuning as discussed in
  Sec.~\ref{sec_tuning}, this does not affect the energy scales of
  different phases (fixed points)\cite{Mitchell2014} nor does it
  affect thermodynamical physical quantities of the model of
  interest.


\section{Methods}
\label{sec_Methods}

The Hamiltonian of quantum impurity models has the form
\begin{equation}
\label{eq:H}
  \hat{H}=
  \hat{H}_{\text{imp}}+\hat{H}_{\text{cpl}}(\{\hat{f}_{0\nu}\})
 +\hat{H}_{\text{bath}}\;.
\end{equation}
It describes an interacting `impurity' subsystem,
$\hat{H}_{\text{imp}}$, coupled by
$\hat{H}_{\text{cpl}}(\{\hat{f}_{0\nu}\})$ to a bath of
non interacting conduction electrons,
\begin{equation}
\label{eq:Hbath}
  \hat{H}_{\text{bath}} =
  \sum_{\nu=1}^{\Nflavor}\sum_{k} 
     \varepsilon_{k\nu}^{\phantom{\dagger}} \hat{c}_{k\nu}^{\dagger}
     \hat{c}_{k\nu}^{\phantom{\dagger}} \;,
\end{equation}
where $\nu=1,\ldots,\Nflavor$ labels the $\Nflavor$ distinct
electron flavors, and $\hat{c}^{\dagger}_{k\nu}$ creates an
electron with a given flavor $\nu$ and momentum $k$ at
energy $\epsilon_{k\nu}\in[-D_{\nu},D_{\nu}]$.  The impurity
is taken to be located at real-space site $\bold{r}={0}$,
and coupled to local bath sites
$\hat{f}_{0\nu}=V^{-1}_{\nu}\sum_k V_{k\nu} \hat{c}_{k\nu}$,
with the normalization factor $|V_{\nu}|^2=\sum_k
|V_{k\nu}|^2$.  The density of bath states with flavor $\nu$
at the impurity position is then given by
$\rho^{\phantom{\rm{d}}}_{\nu}(\omega)=\sum_k
|V_{k\nu}/V_{\nu}|^2 \delta(\omega-\epsilon_{k\nu})$,
defined inside a band of half-width $D_{\nu}$. We assume
constant (momentum-independent) couplings for which the
density of bath states simplifies to a box function,
$\rho^{\phantom{\rm{d}}}_{\nu}(\omega) =
\Theta(\omega-|\epsilon|)/(2D_{\nu})$.  When $\Nchan$
channels of spinful conduction electrons are involved,
$\nu\equiv (\alpha,\sigma)$, where $\alpha \in
\{1,\ldots,\Nchan\}$ labels channels and $\sigma \in
\{\uparrow, \downarrow\}$ labels spins.


\subsection{Standard Wilson chains}
\label{sec:standardWilsonchains}

Within sNRG, $\hat{H}_{\text{bath}}$ is discretized and
mapped onto a 1D tight-binding Wilson chain,\cite{wilson}
consisting of $\Nflavor$ identical `subchains', one for each
flavor. The subchains are constructed as follows: first,
each band $\rho^{\phantom{\rm{d}}}_{\nu}(\omega)$ is divided
up into energy intervals with exponentially reducing width.
The discretization points are given by,
\begin{equation}
\label{eq:discpt}
  \epsilon_{n \nu }^{\pm}(z) = \begin{cases} \pm D_{\nu} \qquad & 
  n=0 \;,\\ \pm D_{\nu} \cdot \Lambda^{-n + z_{\nu}} \qquad & 
  n=1,2,... \;, \end{cases}
\end{equation}
where $\Lambda>1$ is a dimensionless discretization
parameter, and $z_{\nu}\in[0,1[$ (defined modulo 1) is a
continuous `twist' parameter that shifts the discretization
points. Conventionally, the twist parameter is applied
symmetrically to all electronic flavors by choosing
$z_{\nu}\equiv z$. If desired, results of $N_z$ separate NRG
runs with uniformly distributed 
$z$ can be averaged 
to remove
certain discretization artifacts.\cite{Oliveira91,ztrick}

A discretized version of the continuous spectrum
$\rho^{\phantom{\rm{d}}}_{\nu}(\omega)$ is obtained by
replacing the electron density in each interval by a single
pole of the same total weight,
\begin{equation}
\label{eq:discspec}
   \rho^{\text{disc}}_{\nu}(\omega,z)
 = \sum_{n=0}^{\infty}\sum_{\lambda=\pm}
   \gamma^{\lambda}_{n \nu}(z) \delta(\omega-\xi^{\lambda}_{n \nu}(z))  \;.
\end{equation}
where ${\gamma^{\lambda}_{n \nu}(z) = \int_{\epsilon^{\lambda}_{n+1,
\nu}(z)}^{\epsilon^{\lambda}_{n \nu}(z)}
\rm{d}\omega~\rho^{\phantom{\rm{d}}}_{\nu}(\omega)}$ gives
the pole weights.  The pole positions, $\xi^{\lambda}_{n
\nu}(z)$, are determined from a differential equation
introduced in Ref.~\onlinecite{zitko2009}, which is based on
the condition that the original (continuous) bath density of
states is reproduced exactly in the limit $N_z\rightarrow
\infty$ after $z$-averaging, $\rho_{\nu}(\omega) =
\int_0^1\mathrm{d}z~ \rho^{\text{disc}}_{\nu}(\omega,z)$.
For constant density of states, we use
\begin{subequations}
\label{eq:discspecparams}
\begin{align}
   \gamma^{\lambda}_{n \nu}(z) &= D_\nu  
    \begin{cases} 
    1-\Lambda^{z-1}  \qquad & 
   n=0\\ 
   \left( 1-\frac{1}{\Lambda}\right)\Lambda^{-n+z} \qquad & 
   n=1,2,...
   \end{cases}
   \;, \\ 
   \xi^{\lambda}_{n \nu}(z)&=
   \lambda  \frac{\gamma^{\lambda}_{n \nu}(z)} {\ln \Lambda}
    \begin{cases} 
    +z  \qquad & 
   n=0\\ 
   1 
   \qquad & 
   n=1,2,...
   \end{cases}\;.
\label{eq:discspecparams-b}
\end{align}
\end{subequations}
The Wilson subchain for flavor $\nu$ is defined
uniquely\cite{wilson} as the semi-infinite 1D tight-binding
chain that reproduces the discretized density of states
$\rho^{\text{disc}}_{\nu}(\omega)$ at the terminal site.
The discretized bath is represented by the sum of all
$\Nflavor$ Wilson subchains, which together form the `full'
Wilson chain, with Hamiltonian
\begin{equation} 
\label{eq:Hbath_disc}
  \hat{H}_{\text{bath}}^{\text{disc}} = \sum_{\nu=1}^{\Nflavor}
   \sum_{n=0}^{\infty} \left [ \left ( t^{\phantom{\dagger}}_{n \nu}
   f^{\dagger}_{n,\nu}f^{\phantom{\dagger}}_{n+1, \nu} + \text{H.c.}
  \right) + 
  \varepsilon^{\phantom{\dagger}}_{n \nu}
  f^{\dagger}_{n,\nu}f^{\phantom{\dagger}}_{n,\nu} \right] \;,
\end{equation}
The Wilson chain coefficients $t_{n \nu}$ and
$\varepsilon_{n \nu}$ are obtained in practice by Lanczos
tridiagonalization\cite{nrg:rev} [in contrast to the index
$k$ in \Eq{eq:Hbath}, $n$ refers to sites of the Wilson
chain].

Importantly, due to the logarithmic discretization, the hopping matrix
elements decay exponentially along each subchain,\cite{wilson}
\begin{equation}
\label{eq:wc}
  t_{n \nu} / D_{\nu} \sim 
  \Lambda^{z_{\nu}-n/2}\;,
\end{equation}
for $n\gg 1$, and as such depend on NRG discretization
parameters $\Lambda$ and $z_{\nu}$. For equal bandwidths
$D_{\nu}\equiv D$ and constant $z_{\nu}\equiv z$, there is
an energy-scale separation between sites with different $n$,
\begin{equation}
\label{eq:scalesep_nrg}
  t_{ n+1, \nu}/t_{n \nu} ~\overset{\sNRG}{\sim}~ \Lambda^{-1/2} \;.
\end{equation}
However, since the subchains are identical for sNRG, there
is no scale separation between different flavors with the
same site index $n$. Together, these flavors form
`supersite' $n$ of the full Wilson chain: they all have the
\textit{same} characteristic energy scale 
\begin{align}
\label{eq:define-omega_n}
  \w_n=a\Lambda^{-n/2} \, , 
\end{align}
(the constant $a$ is chosen such that the rescaled hoppings
$t_{n-1}/ \w_n\rightarrow 1$ as $n\rightarrow \infty$). As a
consequence, all $\Nflavor$ subsites of supersite $n$ must
be treated equivalently in a single step in sNRG.

The discretized model Hamiltonian in Eqs.~(\ref{eq:H}) and
(\ref{eq:Hbath_disc}) is diagonalized iteratively,\cite{wilson}
starting at the impurity and working down the chain in sNRG
by adding an entire supersite at each iteration $n$. The
energy-scale separation embodied by Eq.~(\ref{eq:scalesep_nrg})
justifies truncation at each step: the lowest $\Nkept$
states are kept, forming a Wilson `supershell', and the
remaining $\Nkept \ttimes (\dbase^{\Nflavor}-1)$ states are
discarded.  If the eigenenergies $E_n$ of supershell $n$ are
given in units of $\w_n$ (`rescaled units'), the typical
level spacing of the lowest-lying levels is of order 1.


\subsection{Interleaved  Wilson chains}
\label{sec:interleavedWilsonchains}

\begin{figure}
\centering
\includegraphics[width=1\linewidth, trim=0mm 19mm 0mm 0mm]{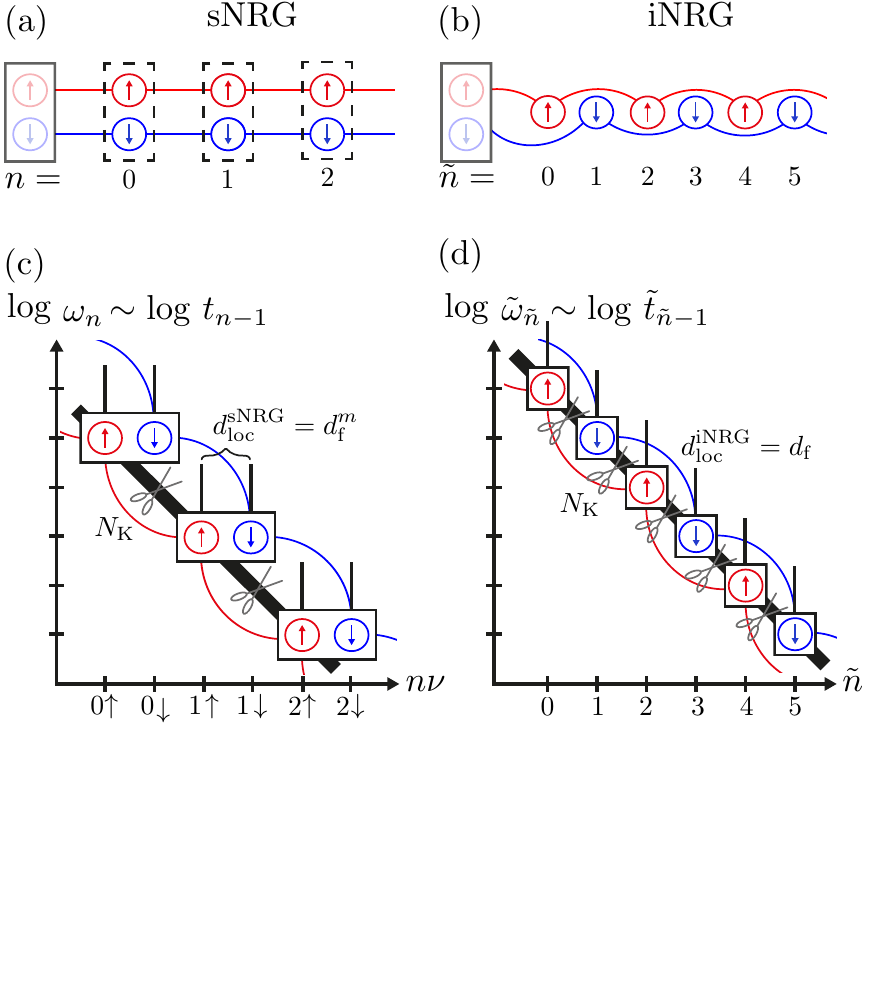} 
\caption{(Color online) 
  Schematic illustration of standard  (left) and interleaved
  (right) Wilson chains, for a spinful single-channel model
  ($\Nflavor=2$). 
  (a) In sNRG, subsites for spin up (red) and spin down
  (blue) are grouped into supersites (indicated by dashed
  boxes), which are connected by nearest-neighbor hopping
  (thin lines).
  (b) In iNRG, subsites are interleaved in linear fashion
  and hopping occurs between next-nearest neighbors.  $n$
  labels sNRG supersites, while $\tilde{n}$ labels iNRG
  subsites.
  (c) and(d) Depictions of the MPS-structure used for sNRG and
  iNRG: boxes represent MPS tensors, vertical thin legs
  represent local state spaces of dimension $\dbase$, and
  thick diagonal lines represent the state spaces obtained
  after diagonalizing a Wilson shell and discarding all but
  the lowest $\Nkept$ states (the truncation process is
  indicated by scissors).  The matrix size to be
  diagonalized is reduced from $\Ntot^\sNRG = \Nkept \ttimes
  \dbase^{\Nflavor}$ in sNRG to $\Ntot^\iNRG = \Nkept
  \ttimes \dbase$ in iNRG.  The vertical positions of the
  boxes reflect, on a logarithmic scale, the characteristic
  energies $\omega_n$ (sNRG) and $\tilde \omega_{\tilde n}$
  (iNRG) of each shell. The additional energy-scale
  separation within each supershell justifies the additional
  truncations in iNRG. 
\label{fig:schematic}}
\end{figure}

We now turn to the iNRG method, introduced in
Ref.~\onlinecite{Mitchell2014}. Its key idea is to modify
the discretization scheme in such a way that energy-scale
separation is achieved between all subsites associated with
the same supersite, as well as between different supersites.
The subsites from different subchains can then be
interleaved in a linear sequence, labeled by $\tilde{n}
\equiv (n,\nu)= m\,n + (\nu-1) = 0,1,2,\ldots$ to form a
single `interleaved' Wilson chain, $\Nflavor$ times longer
than the corresponding standard Wilson chain [compare
\Figs{fig:schematic}(a) and \Figs{fig:schematic}(b)].  The hopping matrix
element $\tildet_{\tilde n} = \tildet_{(n,\nu)}$ describes
hopping between subsites of the same flavor $\nu$ in
adjacent supersites $n$ and $n+1$. For $m>1$, there is  thus
no `nearest-neighbor' hopping on the Wilson chain as in
sNRG. Importantly, $\tildet_{\tilde n}$ progressively
decreases as $\tilde{n}$ increases. To ensure a net rate of
decrease equivalent to that of a standard Wilson chain going
from one supersite to the next [see
Eq.~\eqref{eq:scalesep_nrg}], we have $\tildet_{\tilde
n+\Nflavor}/\tildet_{\tilde n} \propto \Lambda^{-1/2}$.
Moreover, to achieve \textit{uniform} energy-scale
separation along the interleaved chain, this decrease should
occur uniformly from one subsite to the next [see
Fig.~\ref{fig:schematic}(d)], with $\tildet_{\tilde
n+1}/\tildet_{\tilde n} \propto \Lambda^{-1/(2\Nflavor)}$.
By contrast, sNRG amounts to keeping $\tildet_{\tilde n}$
constant for all $\Nflavor$ subsites associated with the
same supersite [see Fig.~\ref{fig:schematic}(c)].  The above
behavior of $\tildet_{\tilde n}$ can be achieved by choosing
the twist parameter $z_{\nu}$ differently for each
conduction electron flavor $\nu$, namely
$z_{\nu+1}=z_{\nu}-1/(2\Nflavor)$. This leads to
\begin{equation}
\label{eq:scalesep_inrg}
   \frac{\tildet_{\tilde{n}+1}}{\tildet_{\tilde{n}}} = 
   \frac{\tildet_{n,\nu+1 }}{\tildet_{n \nu}}
   ~\overset{\iNRG}{\sim}~ \Lambda^{-1/(2\Nflavor)} 
   \equiv \tilde \Lambda^{-1/2} \;,
\end{equation}
with $\tildet_{n,\Nflavor+1 } = \tildet_{n+1,1}$.
Evidently, the effective discretization parameter for iNRG
is smaller than for sNRG, namely $\tilde{\Lambda} \equiv
\Lambda^{1/m}$, thus generating scale separation from
subsite to subsite within a supersite. We  choose $z_m=z$,
such that $\tildet_{(n,m)} = t_n$, i.e.,\ the iNRG hopping
matrix element of the last ($\nu=m$) subsite of supersite
$n$ is identical to the sNRG hopping matrix element for that
supersite. Correspondingly, the characteristic energy scale
for subsite $\tilde{n}$ of the interleaved Wilson chain is
now
\begin{align}
      \label{eq:define-tilde-omega_n}
\tilde{\w}_{\tilde n}=\tilde a\tilde{\Lambda}^{-\tilde{n}/2}
= a \tilde \Lambda^{-(\Nflavor(n-1) + \nu)/2}  \, ,
\end{align}
where the requirement $\tilde{\omega}_{(n,m)} = \w_n$ (which
follows from $\tildet_{(n,m)} = t_n$) fixes the prefactor as
$\tilde a = a \tilde \Lambda^{(m-1)/2}$.  

Scale separation \emph{within} a given Wilson shell $n$ is
exploited in iNRG by performing a truncation after the
addition of each new subsite (rather than only after an
entire supersite of $\Nflavor$ subsites has been added, as
in \sNRG).  With a local state space of
$\dloc^{\iNRG}=\dbase $, at each step $\Nkept$ states are
kept (forming a Wilson `subshell'), and $\Nkept \ttimes\
(\dbase-1)$ ($=\Nkept$ for $\dbase=2$) states are discarded.

If the eigenenergies $\tilde E_{\tilde n}$ of iNRG subshell
$\tilde n$ are measured in rescaled units of
$\tilde{\w}_{\tilde n}$, the spacing of the lowest-lying
levels is again of order 1.  In absolute units, however, the
level spacing in iNRG scales as the $m$-th root compared
with sNRG, because the $m$ subsites are added asymmetrically
(one by one with different hopping matrix elements) in iNRG,
implying $m$ times more iteration steps that lift level
degeneracies. iNRG therefore constitutes a more fine-grained
RG scheme, as illustrated in Fig.~\ref{fig_Etrunc} (compare
the faint red and blue lines).

\subsection{Truncation energy}
\label{sec:Etrunc-1}

In practice,
the value of $\Nkept$ needed
to reach a specified degree of accuracy depends sensitively
on the specific physical model Hamiltonian, discretization
scheme, and energy regime.  
This type of dependence of the accuracy on various details can be circumvented by using
an \emph{energy-based} truncation
strategy,\cite{Weichselbaum2011}
which we also adopt in this paper:
for a given NRG
calculation, we specify a fixed, dimensionless truncation
energy, to be called $\Etrunc^\sNRG$ or $\Etrunc^\iNRG$, and
keep only those states whose absolute (not rescaled)
energies lie below $\text{E}_{\text{abs-trunc}}^{\sNRG}
=\Etrunc^\sNRG \ttimes \w_n$ at iteration $n$ of an sNRG
calculation, or below $\text{E}_{\text{abs-trunc}}^{\iNRG}=
\Etrunc^\iNRG \ttimes \tilde{\w}_{\tilde{n}}$ at iteration
$\tilde{n}$ of an iNRG calculation.  

Using this energy-based truncation scheme, $\Nkept$ becomes
a \emph{dynamical} parameter that changes from iteration to
iteration in a given NRG run, in a way that depends on
$\Etrunc$, $\Lambda$, and details of the particular model
under consideration (Fig.~\ref{fig_Paper_interleaved_5}
below shows an example of the resulting $\Nkept$ values as a
function of iteration number $n$).

When, in our numerical analysis below, we cite sNRG values
for the number of states $\Nkept$ and $\Ntot$ (or for the
corresponding number of symmetry multiplets, $\Nkeptmult$
and $\Ntotmult$), these will refer to the geometric average
over adjacent even and odd sNRG supershells chosen around a
specified energy $E_\mathrm{ref}$ deep in the low-energy
regime, where $E_\mathrm{ref} \ll T_K$.  Similarly, the
corresponding iNRG values refer to a geometric average over
all iNRG subshells associated with both even and odd
supershells near $E_\mathrm{ref}$. 

In general, sNRG calculations performed for the same choice
of $\Etrunc^\sNRG$ yield results of comparable accuracy and
degree of convergence, which are to a large extent
independent of the specific model and discretization
settings being considered.  We have  confirmed this
expectation for the models studied in this paper, as
discussed in detail in Sec.~\ref{sec_Results} below.  For
sNRG, the truncation energy is therefore the key quantity
controlling accuracy and convergence.  

In fact, we find that this is true also for iNRG. Moreover,
we find that sNRG and iNRG calculations yield results with
comparable accuracy and convergence properties, provided
that their truncation energies are related in such a manner
that the resulting $\Nkept^\iNRG $ and $\Nkept^\sNRG$ values
are equal `on average', i.e.,\ after geometrically averaging
over all subsites in a neighboring pair of even and odd
supersites.  We find empirically that this is achieved by
choosing
\begin{eqnarray}
\label{eq:Ectrunc}
  E_{\text{trunc}}^{\iNRG} = 
  E_{\text{trunc}}^{\sNRG} \cdot
 \Lambda^{\tfrac{\Nflavor-1}{4\Nflavor}} \, ,
\end{eqnarray} 
which implies that the parameter $\Etrunc^\iNRG$ is larger than the
parameter $\Etrunc^\sNRG$. Nevertheless, the phrase `equivalent
settings'  includes this choice.  By contrast, the simpler
choice $\Etrunc^\iNRG = \Etrunc^\sNRG$ leads to a smaller average
$\Nkept^\iNRG$ than $\Nkept^\sNRG$.  

In Appendix~\ref{appendix-truncation},
we present a heuristic justification for
Eq.~(\ref{eq:Ectrunc}).
In Sec.~\ref{sec_Results}, sNRG and iNRG results
demonstrate explicitly that the choice of
Eq.~(\ref{eq:Ectrunc}) leads to the desired equivalence of
the number of kept states, accuracy, and convergence.
In the rest of this paper,
we will specify truncation energies in relation to the usual sNRG
value $\Etrunc \equiv \Etrunc^\sNRG$, taking it to be understood that
the corresponding $\Etrunc^\iNRG$ is given by Eq.~\eqref{eq:Ectrunc}.


\subsection{Discarded weight}
\label{sec:discweight}

The convergence of sNRG and iNRG calculations, with a given
truncation threshold $\Etrunc$ and discretization parameter
$\Lambda$, can be analyzed for each model in terms of the
estimated discarded weight,\cite{Weichselbaum2011}
$\delta\rho_{\rm disc}$.  As with the density matrix
renormalization group (DMRG), the decay of the eigenspectrum
of site-specific reduced density matrices, built from the
ground state space of later iterations, can be used as a
quantitative  (\textit{a posteriori}) measure of the
convergence as proposed in
Ref.~\onlinecite{Weichselbaum2011}.  However, in contrast to
Ref.~\onlinecite{Weichselbaum2011}, where only the SIAM with
$\Lambda=2$ was investigated, we wish to compare NRG
calculations performed using a range of different
(effective) discretization parameters $\Lambda$ (or
$\tilde{\Lambda}$) in different models. 	Since the
truncation in NRG is decided on the basis of an
\textit{energy} threshold, in this context it is more
natural to quantify the contributions of high-lying
\textit{energy eigenstates} to reduced density matrices,
rather than analyzing the eigenspectrum of the reduced
density matrices as in Ref.~\onlinecite{Weichselbaum2011}.
The details of our modified approach are presented in 
Appendix~\ref{appendix-discarded}.

By examining the decay of the discarded weight $\delta
\rho_{\rm disc}$ with increasing $\Etrunc$, and observing
the corresponding convergence of physical quantities, we
have found that calculations can be considered converged
when $\delta\rho_{\rm disc}<10^{-6}$.
An important advantage of defining the discarded weight in
terms of the energy eigenbasis is that $\delta \rho_{\rm
disc}$, evaluated at fixed $\Etrunc$, is rather insensitive
to changing the discretization parameter $\Lambda$. 

The discarded weight analysis is particularly important in
benchmarking the iNRG, because the interleaving approach appears
to weaken the energy scale separation ($\tilde{\Lambda} <
\Lambda$). One might then expect\cite{Mitchell2014} that a
larger bare $\Lambda$ would be required in iNRG compared
with sNRG to achieve convergence with the same discarded
weight. However, our detailed study of discarded weights in
Sec.~\ref{sec_Results} in fact reveals the \emph{same}
degree of convergence for iNRG and sNRG when the same
$\Lambda$ and $\Etrunc$ are used.


\subsection{Numerical implementation}

Both sNRG and iNRG can be formulated within the framework of
matrix product states (MPS), which allows for a systematic
and efficient numerical implementation.  Here we employ the
unified tensor representation of the QSpace approach
introduced in Ref.~\onlinecite{Weichselbaum2012}, in which
Abelian and non-Abelian symmetries can be implemented on a
generic level. The state space is labeled in terms of the
symmetry eigenbasis, and the Wigner-Eckart theorem is used
to determine the matrix representation of irreducible
operator sets. Based on this, every (rank-3) tensor object
relevant to NRG calculations splits into a tensor product of
two objects that have identical data structures within the
QSpace approach, operating respectively on the reduced
multiplet space and the Clebsch-Gordan coefficient space.
Matrix diagonalization, for example, is then only performed
in the reduced multiplet space, resulting in an enormous
gain of numerical efficiency.

All correlation functions presented in
Sec.~\ref{sec_Results} are calculated with the
full-density-matrix (fdm-)NRG approach of
Ref.~\onlinecite{Weichselbaum2007}. It is established on a
complete basis set,\cite{asbasisprl,*asbasisprb} constructed
from the discarded states of all NRG iterations.
Since iNRG also produces a matrix-product-state
similar to sNRG, from the point of view of
fdm-NRG, iNRG cannot be distinguished from sNRG.
Therefore the intrinsic multi-shell approach
of fdm-NRG to finite temperature has the major advantage
here that the sub-shell structure of iNRG poses no  complications 
and is automatically taken care of.
Spectral functions for the discretized model then
are given from the
Lehmann representation as a sum of poles, and can be
calculated accurately at zero or arbitrary finite
temperature.  Continuous spectra are obtained by broadening
the discrete data with a standard log-Gaussian kernel of
frequency-dependent width.  \cite{nrg:rev,Weichselbaum2007}


\section{Models}
\label{sec_Models}

In this paper, we study three representative models with
$\Nchan=1,2,$ and $3$ spinful conduction electron channels.
In Sec.~\ref{sec_Results}, iNRG and sNRG are used to solve
these models; the accuracy and efficiency of the two methods
are then compared.  Here we study models with rather high
symmetries; sNRG calculations can exploit either the full
symmetries of the model, or lower symmetries if desired for
comparison with iNRG.  We therefore assume symmetry between
the bands in the following, with half-bandwidth $D_{\nu}
\equiv D =1$ independent of $\nu$. This also sets the
half-bandwidth as the unit of energy.


\subsection{Single impurity Anderson Model ($\Nchan=1)$}
\label{sec_SIAM}

The single impurity Anderson model\cite{hewson} (SIAM)
describes a single correlated quantum level,
\begin{equation}
\label{eq:SIAM}
   \hat{H}_{\rm imp}^{\text{SIAM}} =  \sum_{\nu}\varepsilon_{d \nu} 
   \hat{d}^{\dagger}_{\nu}\hat{d}^{\phantom{\dagger}}_{\nu} + 
   U \hat{d}^{\dagger}_{\uparrow}\hat{d}^{\phantom{\dagger}}_{\uparrow}
   \hat{d}^{\dagger}_{\downarrow}\hat{d}^{\phantom{\dagger}}_{\downarrow}\;,
\end{equation}
tunnel-coupled to a single spinful channel of conduction
electrons $\hat{H}_{\text{bath}}$ (Eq.~\ref{eq:Hbath} with $\Nflavor=2$),
via,
\begin{eqnarray}
   \hat{H}_{\text{cpl}}(\{\hat{f}_{0\nu}\}) &=
   \sum_{k\nu}\left(
      V_{k\nu}\hat{d}^{\dagger}_{\nu}\hat{c}^{\phantom{\dagger}}_{k\nu}
    +\text{H.c.}
   \right )\\
\label{eq:cpl}
  &\equiv \sqrt{\frac{2D\Gamma}{\pi}}
   \sum_{\nu}(\hat{d}^{\dagger}_{\nu}\hat{f}_{0\nu}+{\rm H.c.})\;,
\end{eqnarray}
where $\nu\equiv\sigma\in\{\uparrow,\downarrow\} = \{+,-\}$.
Here $\hat{d}^{\dagger}_{ \nu}$ creates an electron of
flavor $\nu$ on the impurity, with energy $\varepsilon_{d
\sigma}= \varepsilon_d + \sigma h/2$ in a Zeeman field $h$.
For constant, flavor-independent couplings, $V_{k\nu}$, the
hybridization strength is given by
$\Gamma_{\nu}(\varepsilon)=\pi |V_{\nu}|^2
\rho_{\nu}(\varepsilon)\equiv\Gamma \Theta(D-|\varepsilon|)$
within a band of half-width $D\equiv 1$.

The SIAM possesses an SU(2) spin symmetry for $h=0$, to be
denoted by $\text{SU(2)}_\spin$, which reduces to
$\text{U(1)}_\spin$ for $h \neq 0$. Moreover, at
particle-hole symmetry, $\varepsilon_{d} = -U/2$, the SIAM
possesses an SU(2) symmetry involving transformations
between particles and holes, to be called
$\text{SU(2)}_\charge$. This reduces to
$\text{U(1)}_\charge$ for $\varepsilon_{d} \neq -U/2$.
Depending on the symmetries allowed by the choice of model
parameters, sNRG can exploit any combination of these spin
and charge symmetries. In this paper we set $h\! =\! 0 $ and
$\varepsilon_{d} = -U/2$, and employ either
$\text{U(1)}_\spin \ttimes \text{U(1)}_\charge $ or
$\text{SU(2)}_\spin \ttimes \text{SU(2)}_\charge $
symmetries.

Within iNRG, we can interleave Wilson chains for the
$\nu=~\uparrow$ and $\downarrow$ conduction electrons
species, discretizing these separately for a given $\Lambda$
using two different $z$-shifts,
$z_{\uparrow}=z+\tfrac{1}{4}$ and $z_{\downarrow}=z$.  Since
this `spin-interleaved' scheme (spin-iNRG) artificially
breaks the bare symmetry between spin up and down, it
reduces the $\text{SU}(2)_\spin$ symmetry to
$\textrm{U(1)}_\spin$.  Furthermore, $\text{SU(2)}_\charge$
is reduced to $\textrm{U(1)}_\charge$ in spin-iNRG, as the
irreducible operator set for $\text{SU(2)}_\charge$ mixes
spin components, and therefore cannot be defined within the
state space of a single fixed-spin subsite.  Consequently,
spin-iNRG studies of the SIAM can employ  $\text{U(1)}_\spin
\ttimes \text{U(1)}_\charge $ symmetries only.


\subsection{Two-channel Kondo model ($\Nchan=2$)}
\label{sec_2ck}

The two-channel Kondo model (2CKM),\cite{Nozieres} features
a single spin-$\tfrac{1}{2}$ impurity with spin
$\hat{\bold{S}}_{\frac12}$ coupled by antiferromagnetic
Heisenberg exchange to two spinful conduction electron
channels (Eq.~\ref{eq:Hbath} with $\Nflavor=4$),
\begin{equation}
\label{eq:2ckm}
   \hat{H}_0^{\text{2CKM}} = 
   \sum_{\alpha} J_{\alpha} \hat{\bold{S}}_{\frac12} \cdot
   \hat{\bold{s}}_{\alpha} 
   + h S^z_{\frac12} \;,
\end{equation}
where $\hat{\bold{s}}_{\alpha} = \sum_{\sigma\sigma'}
\hat{f}^{\dagger}_{0\alpha  \sigma}\tfrac{\vec{\sigma}_{\sigma\sigma'}}{2}
\hat{f}^{\phantom{\dagger}}_{0\alpha\sigma'}$
is the conduction electron spin density at the impurity in channel
$\alpha=1,2$ (and $\vec{\sigma}$ is a vector of Pauli matrices).

In the spin sector, the 2CKM possesses an
$\text{SU(2)}_\spin$ symmetry for $h=0$, and an
$\text{U(1)}_\spin$ symmetry for $h \neq 0$.  In the case of
particle-hole and channel symmetry  ($J_1=J_2$), the
$\Nflavor=4$ flavors possess the enlarged symplectic
symmetry $\text{Sp}(4)_\chargechannel$.  This reduces to
$[\text{SU}(2)_\parthole]^2$ if channel symmetry is broken
($J_1\ne J_2$), and further to $[\text{U}(1)_\parthole]^2$
if particle-hole symmetry is broken (not considered here).
Depending on the symmetries allowed by the choice of model
parameters, sNRG can exploit any combination of these spin
and charge symmetries.  We will here set $h =0 $ and employ
either the $\text{U(1)}_\spin \!\times\!
[\text{U(1)}_\charge]^2 $, $\text{SU(2)}_\spin \!\times\!
[\text{SU}(2)_\charge]^2 $, or $\text{SU(2)}_\spin \!\times\!
\text{Sp}(4)_\chargechannel $ symmetries.

Within iNRG, the four electron flavors can be interleaved in
several different ways. For example, using spin-iNRG (as
described above, with $z_{\alpha,\uparrow}=z+\tfrac{1}{4}$
and $z_{\alpha,\downarrow}=z$), the spin symmetry is reduced
to $\text{U(1)}_\spin$. Although $z_{1,\sigma} =
z_{2,\sigma}$, this can only be combined with
$\text{U(1)}_\charge$  symmetries in the particle sector.

Alternatively, one can interleave the spinful $\alpha=1,2$
channels, discretizing them separately using
$z_{\alpha=1,\sigma}=z+\tfrac{1}{4}$ and
$z_{\alpha=2,\sigma}=z$ (but
$z_{\alpha,\uparrow}=z_{\alpha,\downarrow}$). This
`channel-interleaved' scheme (channel-iNRG) breaks the
symmetry between channel 1 and 2 (even if $J_1=J_2$) and
hence the full $\text{Sp}(4)_{\chargechannel}$ symmetry is
broken. However, $[\text{SU}(2)_\parthole]^2$ symmetry can
still be exploited, in combination with either
$\text{SU(2)}_\spin$ or $\text{U(1)}_\spin$.

In the most asymmetric case, all four electron flavors of
the 2CKM are interleaved, using
$z_{1\uparrow}=z+\tfrac{3}{8}$,
$z_{1,\downarrow}=z+\tfrac{2}{8}$,
$z_{2,\uparrow}=z+\tfrac{1}{8}$, $z_{2,\downarrow}=z$.  The
maximum symmetry consistent with this `flavor-interleaved'
scheme (flavor-iNRG) is $\text{U}(1)_\spin\!\times\!
[\text{U}(1)_\parthole]^2$.  In this paper, our iNRG studies
of the 2CKM will employ the latter flavor-iNRG scheme, and
also channel-iNRG with $\text{SU}(2)_\spin\!\times\!
[\text{SU}(2)_\parthole]^2$ symmetry.

The point $J_1\rho_1(0)=J_2\rho_2(0)$ is a critical point of
the 2CKM, characterized by a frustration of screening that
gives rise to fragile non-Fermi liquid
physics.\cite{Nozieres,Affleck1992} Any finite channel
anisotropy $J_1\rho_1(0) \ne J_2\rho_2(0)$ produces a
crossover\cite{Affleck1992,akm:exactNFL,*akm:finiteT} to a
Fermi liquid ground state, corresponding to Kondo strong
coupling between the impurity and channel $\alpha=1$ (or $2$)
for $\left(J_1\rho_1(0)\right)/\left(J_2\rho_2(0)\right)> 1$
(or $<1$). Because the interleaving in iNRG spoils the channel
symmetry [$\rho_1^{\rm disc}(0)\neq\rho_2^{\rm disc}(0)$]
even in the isotropic case $J_1=J_2$, the critical point of
the 2CKM is spuriously destabilized. fine tuning of the
ratio $J_1/J_2\approx 1$ must then be carried out to access
the critical physics.\cite{Mitchell2014} This is discussed
further in Sec.~\ref{sec_tuning}. 


\subsection{Anderson-Hund model ($\Nchan=3$)}
\label{sec_3CAHM}

Finally, we consider the particle-hole symmetric
three-channel Anderson-Hund model (3CAHM) of
Refs.~\onlinecite{Weichselbaum2012,Hanl2013,Hanl2014}. The
isolated `impurity', comprising $\alpha=1,2,3$ orbitals,
each with spin $\sigma=~\uparrow,\downarrow$, is described by
\begin{equation}
\label{eq:3cahm}
   \hat{H}_{\rm imp}^{\text{3CAHM}} =  -J_H \hat{\bold{S}}^2  \;,
\end{equation}
where
$\hat{\bold{S}}=\sum_{\alpha}\sum_{\sigma,\sigma'}
   d^{\dagger}_{\alpha\sigma} \tfrac{\vec{\sigma}_{\sigma\sigma'}}{2}
   d^{\phantom{\dagger}}_{\alpha\sigma'}
$
is the \emph{total} impurity spin. Electrons of different
impurity orbitals interact through the Hund coupling, $J_H$,
in Eq.~\eqref{eq:3cahm}.

Each impurity orbital with flavor $\nu = (\alpha,\sigma)$ is
tunnel-coupled to a conduction electron band of the same
flavor, via Eq.~(\ref{eq:cpl}); overall there are
$\Nflavor=6$ electronic flavors.  The large local state
space $\dloc^{\sNRG}=64$ for the 3CAHM means that iterative
diagonalization in \emph{state} space (rather than multiplet
space) is practically intractable for sNRG. However, this
3CAHM possesses large symmetries that can be optimally
exploited in sNRG: $\text{SU}(2)_\spin$ symmetry in the spin
sector, and $\text{Sp}(6)_\chargechannel$ symmetry in the
particle-hole/channel sector. The 64 states describing a
single Wilson supersite reduce to a mere 4 multiplets in
this case. 

The $\text{Sp}(6)_\chargechannel$ symmetry reduces to
$\text{U(1)}_\charge \!\times\! \text{SU(3)}_\channel$ if
particle-hole symmetry is broken, or to
$[\text{SU(2)}_\charge]^3$ if channel symmetry is broken.
Exploiting one of these three large symmetries is essential
when using sNRG.

For iNRG, one again has several options for interleaving.
We will consider channel-iNRG with
$\text{SU}(2)_\spin\!\times\! [\text{SU}(2)_\parthole]^3$
symmetry, and full flavor-iNRG with
$\text{U}(1)_\spin\!\times\! [\text{U}(1)_\parthole]^3$
symmetry. 

A major advantage of iNRG is that such models can be
solved even when no large symmetries are available [cf.
yellow dashed curve in Fig.~\ref{fig_Paper_interleaved_2}(b)
that shows the spectral function of the 3CAHM calculated
with iNRG and $\text{U}(1)_\spin\!\times\!
[\text{U}(1)_\parthole]^3$ symmetry].

\begin{center}
\begin{figure*}
\includegraphics[width=1\linewidth, trim=6mm 3mm 6mm 0mm]{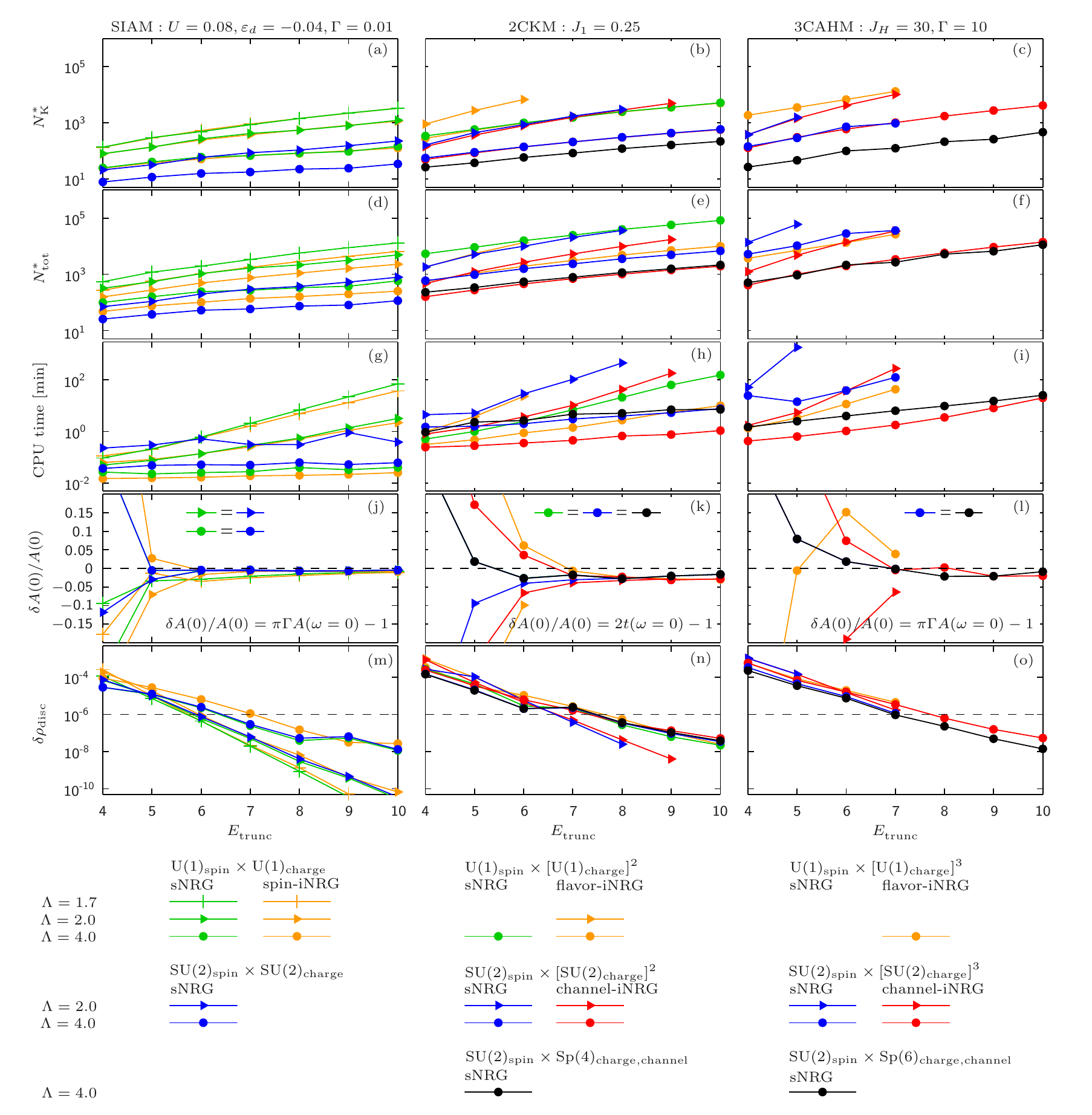}
\caption{(Color online)
  Comparison of sNRG and iNRG for three models: SIAM (left
  column), 2CKM (middle column) and 3CAHM (right column).
  (a-c) The number of kept multiplets, $\Nkeptmult$; (d-f)
  the total number of multiplets generated during an NRG
  step, $\Ntotmult$; (g-i) the total CPU time for one NRG
  run;  (j-l) the relative deviations $\delta A(0) /A(0)$
  of correlation functions at the Fermi energy $\omega=0$
  from their exact values [cf.\ Eqs.~\eqref{eq:deviations}];
  and (m-o) the discarded weight $\delta \rho_{\rm disc}$
  (the horizontal dashed lines indicate the convergence
  threshold).  All quantities (computable with a maximum
  memory of 128 GB) are plotted versus $\Etrunc$ ($\equiv
  \Etrunc^\sNRG$), for $\Lambda = 1.7$ (crosses), 2.0
  (triangles) and 4.0 (circles). Each symmetry setting is
  identified by a particular color in iNRG and sNRG. iNRG
  results have been geometrically averaged over all
  interleaved flavors.  Data for $\Nkeptmult$ and
  $\Ntotmult$ have been geometrically averaged over even and
  odd Wilson shells at an energy scale $E_\mathrm{ref} =
  5\times 10^{-8}D \ll T_K$.  We used $z=0$ in all cases
  except for panels (j-l), where data for $z=0$ and $0.5$
  have been averaged.  An exception is the flavor-iNRG data
  point at $\Etrunc=7$ in panel (l), which was obtained for
  $z=0$ without $z$-averaging ($z=0.5$ exceeded memory
  resources).  In panels (j-l), sNRG results with the same
  $\Lambda$ but different symmetry settings coincide. \\
}
\label{fig_Paper_interleaved_1}
\end{figure*}
\end{center}
%


\section{Results}
\label{sec_Results}

In the following, we present a comprehensive comparison of
iNRG and sNRG for the three models introduced in
Sec.~\ref{sec_Models}. We begin in
Sec.~\ref{sec_results_overview} by summarizing our main
conclusions, referring only briefly to the relevant figures.
We then offer a detailed analysis of the figures to
substantiate our main results in the subsequent sections. In
particular, we compare iNRG and sNRG by examining the number
of kept multiplets in Sec.~\ref{sec_results_Nkept}, the
efficiency of the calculations in
Sec.~\ref{sec_results_efficiency}, and the
accuracy/convergence of the results in
Sec.~\ref{sec_results_accuracy}. The take-home message is
that iNRG offers significant improvements in efficiency
without compromising accuracy and convergence properties.
   

\subsection{Overview}\label{sec_results_overview}
We perform calculations in which, for a given model,
discretization parameter $\Lambda$, and choice of exploited
symmetries, the truncation energies of iNRG and sNRG are
related by \Eq{eq:Ectrunc}. This use of \textit{equivalent
settings} allows for optimal comparability, because it
ensures that, on average, the same number of states are kept
at each iteration in both methods. The number of kept
\emph{multiplets}, $\Nkeptmult$, is therefore also the same
on average -- as demonstrated explicitly in
Figs.~\ref{fig_Paper_interleaved_1}(a-c) and
\ref{fig_Paper_interleaved_5}.

\textit{Number of multiplets:}
$\Nkeptmult$, and thus also $\Nkept$, is found to increase
roughly exponentially with $\Etrunc$ and also with the
number of conduction electron channels $\Nchan$
[Figs.~\ref{fig_Paper_interleaved_1}(a-c) and
\ref{fig_Paper_interleaved_4}].  This scaling is common to
both iNRG and sNRG. It simply reflects the fact that the
number of many-body eigenstates of a gapless system grows
exponentially with energy, with an exponent that increases
linearly with $\Nchan$.  Since we exploit symmetries and
conserved quantities in the calculations, the number of kept
\emph{multiplets} $\Nkeptmult$ is far smaller than the
number of kept \emph{states} $\Nkept$ in both iNRG and sNRG.
When iNRG and sNRG use the same symmetry setting, the total
number of multiplets to be diagonalized at each iteration,
$\Ntot^{*}$, is far smaller for iNRG than sNRG
[Figs.~\ref{fig_Paper_interleaved_1}(d-f)], due to the
intermediate truncations in iNRG.  However, iNRG cannot
always exploit the full model symmetries due to the
interleaving process. As a consequence there can be an
efficiency tradeoff in iNRG: the advantage of a reduced
local state space comes at the cost of fewer symmetries
being available to exploit.  This is shown by
Fig.~\ref{fig_Paper_interleaved_1}(d-f), where $\Ntot^{*}$
for the most efficient iNRG calculation is essentially the
 same as that of the best sNRG calculation (exploiting all
 symmetries) in each case [in fact, $\Ntot^{*}$ is actually
 lower in sNRG for the SIAM in panel (d)].

\textit{Efficiency:}
The total CPU time for a given iNRG calculation is smaller
than that of the corresponding sNRG calculation with
equivalent settings
[Fig.~\ref{fig_Paper_interleaved_1}(g-i)]. In fact, with
$\Lambda = 4$, spin-iNRG for the SIAM, spin and channel-iNRG
for the 2CKM and channel-iNRG calculations for the 3CAHM are
also more efficient than the \emph{best} sNRG calculations
exploiting full symmetries.  Even though $\Ntot^{*}$ is
typically similar or even lower for the best sNRG compared
to the best iNRG calculations, the book-keeping overheads
involved in exploiting symmetries can outweigh the
potential gains of doing so (this is especially pronounced
for smaller $\Etrunc$).  In general, the gain in iNRG
efficiency becomes more significant as the number of
flavors increases.  Importantly, some low-symmetry,
many-band models that are prohibitively expensive for sNRG
can still be tackled with iNRG.

\textit{Accuracy and convergence:} 
Remarkably, these gains in efficiency do not compromise
accuracy and convergence properties. To establish this, we
performed extensive comparisons between iNRG and sNRG using
equivalent settings [see
Figs.~\ref{fig_Paper_interleaved_1}(j)-\ref{fig_Paper_interleaved_1}(l), \ref{fig_Paper_interleaved_1}(m)-\ref{fig_Paper_interleaved_1}(o), and
\ref{fig_Paper_interleaved_2}].

The accuracy of iNRG was established directly, by monitoring
the deviation of calculated physical quantities from certain
exact results.  In particular, we studied the value of the
impurity spectral function (or t matrix) at the Fermi level,
relative to known analytic results
[Figs.~\ref{fig_Paper_interleaved_1}(j)-\ref{fig_Paper_interleaved_1}(l)].  The quality of
the results improves with increasing $\Etrunc$ as expected,
and exact results are reproduced to within a few percent for
$\Etrunc>7$ in both sNRG and iNRG. This conclusion is
further supported by comparisons of the full frequency
dependence of impurity spectral functions in
Fig.~\ref{fig_Paper_interleaved_2}.

Furthermore, our analysis of the discarded weight shows that
both iNRG and sNRG calculations are effectively converged
for $\Etrunc>7$ [Figs.~\ref{fig_Paper_interleaved_1}(m)-\ref{fig_Paper_interleaved_1}(o)].
This demonstrates explicitly that the states discarded at
intermediate steps in iNRG do not contribute appreciably to
low-energy eigenstates at later iterations, thus validating
the more fine-grained RG scheme employed by iNRG.

\textit{Artificially broken symmetries.}
Finally, we examined the tuning protocol employed in iNRG to
restore channel symmetries that are broken artificially by
the interleaved discretization (see
Fig.~\ref{fig_Paper_interleaved_3}). Such channel symmetries
are of course not always relevant perturbations (an example
is the 3CAHM, where the same basic low-energy physics arises
even in the channel-anisotropic case). The worst-case
scenario for iNRG emerges in the vicinity of a quantum
critical point, where channel asymmetries generate a
relevant RG flow to a different fixed
point.\cite{Nozieres,Affleck1992} The classic exemplar is
the 2CKM, whose frustrated critical point occurs precisely
at $J_1\rho_1(0)=J_2\rho_2(0)$. In iNRG, where $\rho_1^{\rm
disc}(0)\neq \rho_2^{\rm disc}(0)$, the ratio $J_1/J_2$ must
be tuned to access this physics, but is found in practice to
deviate from its exact value by only $\sim 1\%$.  We also
show that the critical point can be located exponentially
rapidly in the number of iNRG runs, keeping calculation
overheads to a minimum.


\subsection{Number of kept multiplets}
\label{sec_results_Nkept}
The key difference between sNRG and iNRG is the size of the
local state space -- i.e.\ $\dloc^\sNRG=\dbase^m$ vs.\
$\dloc^\iNRG=\dbase$. However, to compare fairly the
relative efficiency, we must ensure that both calculations
are of comparable accuracy. By choosing the `same'
truncation energies in iNRG and sNRG [via
Eq.~(\ref{eq:Ectrunc})], a comparable number of multiplets
is kept in both calculations, as argued in
Sec.~\ref{sec:Etrunc-1}. Here we present data to
substantiate this. Moreover, the consequence of this choice
is that sNRG and iNRG calculations are of equivalent
accuracy, as demonstrated explicitly below in
Sec.~\ref{sec_results_accuracy}.

\begin{figure}[ht]
\centering
\includegraphics[width=1\linewidth, trim=0mm 0mm 0mm 0mm]{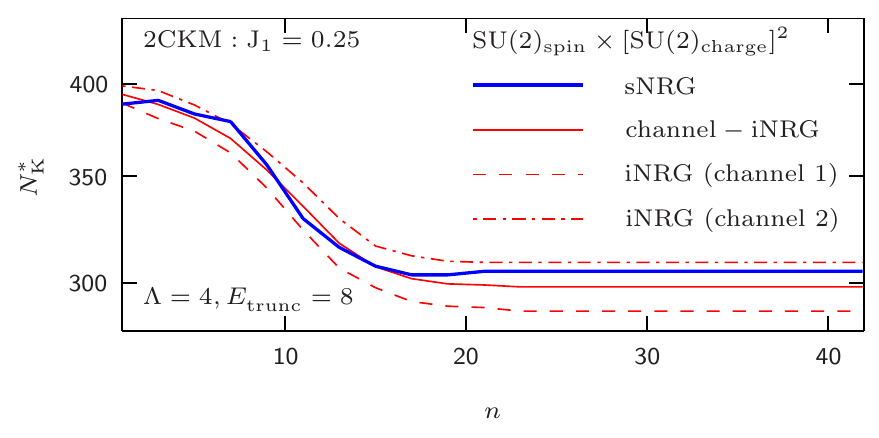} 
\caption{(Color online) 
  Number of kept multiplets $\Nkeptmult$, vs Wilson
  shell index $n$ for the 2CKM within the
  $\text{SU}(2)_\spin\!\times\! [\text{SU}(2)_\parthole]^2$
  symmetry setting. For iNRG (red), the number of multiplets
  kept after adding channel 1 (dashed) or channel 2
  (dash-dotted) are shown separately, as well as their
  geometric average (solid).  sNRG results are shown in blue
  for comparison. All results are geometrically averaged
  over even and odd iterations.
}
\label{fig_Paper_interleaved_5}
\end{figure}

Figures~\ref{fig_Paper_interleaved_1}(a) and \ref{fig_Paper_interleaved_1}(c) show $\Nkeptmult$
obtained for the SIAM, 2CKM, and 3CAHM, with several
different choices of $\Lambda$, and employing various
symmetry settings. In all cases, we find that
$\Nkept^{\ast,\iNRG}$ and $\Nkept^{\ast,\sNRG}$ are
comparable when the same symmetry setting is used. However,
note that the different iNRG subshells contribute unequally
to their geometric average, because the absolute truncation
energy changes from subshell to subshell in iNRG, as
explained in Sec.~\ref{sec:Etrunc-1}.  This is illustrated
in Fig.~\ref{fig_Paper_interleaved_5}, which shows
$\Nkeptmult$ as function of Wilson shell index $n$ for the
2CKM.  For iNRG, the number of multiplets kept after adding
the first channel (red dashed line) is smaller than  the
number of multiplets kept after adding the second channel
(red dash-dotted), but their geometric average (red solid)
is rather similar to the number of kept multiplets in the
corresponding sNRG calculation (blue solid line) for all
$n$. 

Furthermore, \Figs{fig_Paper_interleaved_1}(a) and \ref{fig_Paper_interleaved_1}(c)  confirm
that the total number of kept multiplets $\Nkept^{*}$
depends exponentially on $\Etrunc$, with a growth exponent
that increases with $N_c$ [the slope of the line increases
from \Fig{fig_Paper_interleaved_1}(a) to
\ref{fig_Paper_interleaved_1}(c)]. This behavior is expected
for the many-body eigenstates of a gapless system, whose
number increases exponentially with energy.

Naturally, exploiting larger symmetries means that fewer
multiplets are kept for a given $\Etrunc$. [For example, in
\Fig{fig_Paper_interleaved_1}(c) for $\Lambda = 4$,  the
black circles lie well below the red and blue circles.] This
reduction of the multiplet space arises by splitting off
large Clebsch-Gordan spaces. We also note that smaller
$\Lambda$, which reduces energy-scale separation between
iterations, leads to larger $\Nkeptmult$, and to a faster
increase of $\Nkeptmult$ with $\Etrunc$. [For example, in
\Fig{fig_Paper_interleaved_1}(b), the blue and red triangles
for $\Lambda =2$ lie above the blue and red circles for
$\Lambda = 4$, and rise with a greater slope.]

\begin{figure}[ht]
\centering
\includegraphics[width=1\linewidth, trim=0mm 5mm 0mm 0mm]{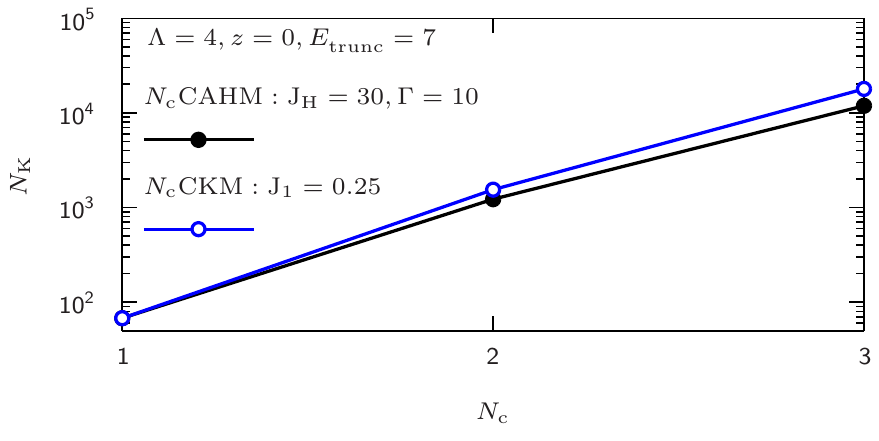}
\caption{(Color online)
  Number of kept \emph{states}, $\Nkept$, at the low-energy
  fixed point of the $\Nchan$CAHM (black) and $\Nchan$CKM
  (blue), showing a roughly exponential increase with the
  number of channels,  $\Nchan$. Results were obtained with
  sNRG and have been geometrically averaged over even and
  odd Wilson shells at energy scale $E_\mathrm{ref} =
  5\times 10^{-8}D \ll T_K$. 
}
\label{fig_Paper_interleaved_4}
\end{figure}

Furthermore, the number of kept \emph{states}, $\Nkept$
(which is independent of the symmetry settings used),
increases roughly exponentially with the number of
conduction electron channels, $\Nflavor$.  This is confirmed
in Fig.~\ref{fig_Paper_interleaved_4}, which shows $\Nkept$
for the multi-channel Kondo model ($\Nchan$-CKM) and the
multi-channel Anderson-Hund model ($\Nchan$-CAHM), with
$\Nchan=1,2,3$ spinful channels [these models are the
generalizations of Eqs.~(\ref{eq:2ckm}) and
(\ref{eq:3cahm}) to the case of $\Nchan$ channels].
Figure \ref{fig_Paper_interleaved_4} also shows that the
description of certain multi-channel fixed points requires
a greater number of kept states than others, reflecting
their relative complexity. For example, the frustrated
non-Fermi liquid fixed points of the $\Nchan$CKM (with
$\Nchan\ge 2$) require a larger $\Nkept$ than the
corresponding Fermi liquid fixed points of the $\Nchan$CAHM
at $\Etrunc=7$.

These results confirm that the exponential scaling of
required computational resources with $\Nflavor$ in both
sNRG and iNRG cannot be avoided -- it simply reflects
elementary state-counting properties for gapless
multi-channel systems.  However, the efficiency of the
calculation for a given model can be substantially improved
by exploiting symmetries in sNRG, or by interleaving flavors
in iNRG, as now discussed.


\subsection{Efficiency}
\label{sec_results_efficiency}

The \emph{total} number of multiplets, $\Ntotmult$,
generated in NRG near the low-energy fixed point of the
three models, is plotted as a function of $\Etrunc$ in
Figs.~\ref{fig_Paper_interleaved_1}(d-f).  As with
$\Nkept^{*}$, the size of $\Ntotmult$ depends on the
particular model under consideration, $\Lambda$, $\Etrunc$,
and the symmetry setting used.  Additionally, we now also
see a dramatic difference between iNRG and sNRG.  When the
same symmetry setting is used, $\Ntotx{\iNRG}$ is far
smaller than $\Ntotx{\sNRG}$, because $\dloc^{\rm{iNRG}}$ is
smaller than $\dloc^{\rm{sNRG}}$.  [For example: the red data
points lie clearly below the blue data points in
\Fig{fig_Paper_interleaved_1}(f) for the 3CAHM and in
\Fig{fig_Paper_interleaved_1}(e) for the 2CKM.] Moreover,
then also the ratio $\Ntotx{\sNRG}/\Ntotx{\iNRG}$ grows
exponentially with the number of interleaved flavors.  For a
given model and symmetry, $\Ntotx{\sNRG}/\Ntotx{\iNRG}$
would therefore be larger for full flavor-iNRG than
channel-iNRG. However, note that in general $\Ntotmult$
itself might be smallest for channel-iNRG, meaning that the
optimal strategy might involve keeping some symmetries at
the expense of interleaving fewer flavors. An example of
this is seen clearly for the 2CKM in
Figs.~\ref{fig_Paper_interleaved_1}(e,f), where red dots lie
below orange dots.

These trends in $\Ntotmult$ are reflected in the CPU time
plotted in Figs.~\ref{fig_Paper_interleaved_1}(g-i), which
is the ultimate measure of calculation efficiency. The total
CPU time for an NRG calculation is generally dominated by
matrix diagonalizations (especially for large $\Etrunc$),
and therefore scales as $\sim (\Ntot)^3$. Since $\Nkept^{*}$
and $\Ntot^{*}$ grow with $\Etrunc$, so too does the CPU
time -- the faster so with smaller $\Lambda$.  For small
$\Etrunc$, however, numerical overheads can also have a
noticeable influence. [Example: in
\Fig{fig_Paper_interleaved_1}(e) for $\Ntotmult$, the green
and orange circle points for $\Lambda=4$ show a separation
that is quite large and approximately constant; by contrast,
\Fig{fig_Paper_interleaved_1}(h) shows an increasing
difference in the CPU time with increasing $\Etrunc$. At
large $\Etrunc$, the difference is essentially attributable
to the difference in $\Ntotmult$ alone. At small $\Etrunc$
the numerical overhead in iNRG can presumably be attributed
to larger Wilson chain lengths.]

The maximum efficiency gain of iNRG over sNRG in terms of
CPU time occurs if no symmetries are used in either iNRG or
sNRG. This gain is then of order $\sim
\dbase^{3(\Nflavor-1)}/\Nflavor$, where the factor of
$1/\Nflavor$ arises because the interleaved Wilson chain is
$\Nflavor$ times longer than the standard Wilson chain.
Similarly, the corresponding gain in terms of memory
resources is given by $\sim \dbase^{2(\Nflavor-1)}$, here
without the factor of $1/\Nflavor$, since memory is required
on the level of a specific NRG iteration rather than for the
whole calculation.  The following table summarizes the
theoretical maximum gain relative to sNRG (in the absence of
symmetries) obtained with channel-iNRG and flavor-iNRG for
models with $\Nchan=1,2,3$:
\begin{center}
\begin{tabular}{|c|c|c|c|}
  \hline
   No.~spinful & No.~channels & Max.~speedup& Max. gain  \\
   channels & interleaved & factor (CPU) & in memory \\
  \hline \hline
	$\Nchan=1$ & $N_{\nu=\sigma}=2$ & $4$ & 4 \\ 
	\hline
  $\Nchan=2$ & $N_{\nu=\alpha}=2$ & $32$  & 16 \\ \cline{2-4}
   & $N_{\nu=\alpha\sigma}=4$ & $128$  & 64 \\ 	
  \hline
  $\Nchan=3$ & $N_{\nu=\alpha}=3$ & $ 1365$  & 256 \\ \cline{2-4}
   & $N_{\nu=\alpha\sigma}=6$ & $5461$  & 1024 \\ 	
  \hline
\end{tabular}
\end{center}

When symmetries are exploited in the calculations, the
efficiency gain for iNRG over sNRG is reduced, relative to
the value cited in the above table, because the local
Hilbert space of each supersite in sNRG (or subsite in iNRG)
is organized into \textit{multiplets} instead of states. The
factor ${\dbase}^{3(\Nflavor-1)}/\Nflavor$, which was based
on a \textit{state}-counting argument, is then effectively
reduced. Note, however, that handling and book-keeping of
Clebsch-Gordan coefficient spaces also introduces a
numerical overhead. For very small $\Etrunc$, this can even
outweigh the efficiency gains of exploiting symmetries.
However, the symmetry gains grow with increasing $\Etrunc$
(which leads to increasingly large block sizes for reduced
matrix elements), and eventually always dominate compared to
book-keeping overheads. [Example: in
\Fig{fig_Paper_interleaved_1}(h) for $\Lambda=4$, the green
circles lie below the blue circles for small $\Etrunc$, but
cross at $\Etrunc \simeq 6$. For large $\Etrunc$, the most
efficient sNRG calculations are those that exploit the
largest symmetries (the black circles start crossing the
blue circles at $\Etrunc=10$).] 

Ultimately, when the same symmetries are used for both
calculations, iNRG clearly requires far smaller CPU time
than sNRG for a given $\Lambda$ and $\Etrunc$ -- see
Figs.~\ref{fig_Paper_interleaved_1}(h,i). This effect
becomes more pronounced with increasing $\Etrunc$. 

The models considered here have high intrinsic symmetries,
which can be more fully exploited in sNRG than iNRG.  The
`best case' scenario for sNRG, in which the \emph{full}
model symmetries are exploited, are shown as the blue data
points in the first column of
Fig.~\ref{fig_Paper_interleaved_1} and as black points in
the second and third columns of
Fig.~\ref{fig_Paper_interleaved_1}. For $\Nchan=1$ [panel
(d)], this optimal sNRG generates a slightly smaller
$\Ntotmult$ than the best corresponding iNRG calculation
(for a given $\Lambda$ and $\Etrunc$). However, when the
number of channels is increased to $\Nchan=2$ or $3$ [panels
(e) and (f)], we find similar $\Ntotmult$ for the best iNRG
calculations (red dots) and the best sNRG calculations
(black dots). Nevertheless, for the range of $\Etrunc$
values used here, the total CPU times [panels (h) and (i)] for
$\Lambda=4$ calculations employing channel-iNRG (red dots)
are still lower than for sNRG (black dots), even when full
symmetries are exploited in sNRG (the difference is
attributable to additional book-keeping costs incurred when
handling large symmetries in sNRG). 

The benefits of exploiting symmetries increase for larger
$\Ntotmult$ and hence $\Etrunc$. As a consequence, we find
that the CPU time increases with  $\Etrunc$ slower for
full-symmetry sNRG than for iNRG.  For example, for $\Nchan
= 3$ and $\Lambda = 4$, in panel (f) for $\Ntotmult$ the
black (sNRG) and red (iNRG) dots are approximately
equivalent, while in panel (i) for the CPU times, the black
dots lie well above the red dots for small $\Etrunc$, but
then rise more slowly with $\Etrunc$, so that both roughly
coincide for $\Etrunc=10$.  Similarly, for $\Nchan =1$ and
$\Lambda = 2$, in panel (g) for $\Ntotmult$ the blue
triangles (sNRG) start above the orange triangles (iNRG) for
small $\Etrunc$, but rise more slowly and end up below the
latter for $\Etrunc \gtrsim 7$, showing that full-symmetry
sNRG can sometimes be the most efficient method. We also
anticipate that full-symmetry sNRG for $\Nchan = 3$ [panel
(i)] would be more efficient than iNRG for $\Etrunc \gtrsim
10$.

Finally, we note that the optimal iNRG calculation does not
necessarily involve interleaving all possible flavors, due
to the tradeoff in lowered symmetries.  Indeed, making
partial use of interleaving and partial use of symmetries
can yield the best results, as seen, for example in panels
(h) and (i) for the CPU times of $\Nchan = 2$ and $3$,
respectively, where the red symbols (channel-iNRG) lie below
the corresponding orange symbols (flavor-iNRG). 
  

\subsection{Accuracy}
\label{sec_results_accuracy}

As highlighted above, the iNRG scheme is more efficient due
to the intermediate truncations along the interleaved Wilson
chain, which results in the smaller local state space
$\dloc=\dbase$ at each step (if all flavors are
interleaved). A key question is whether these intermediate
truncations adversely affect the accuracy of iNRG results.
In the following, we show that, for the same model and same
$\Lambda$, with truncation energies set equal as in
Eq.~(\ref{eq:Ectrunc}), we obtain results with similar
accuracy and convergence properties for both iNRG and sNRG.

\begin{figure}[tb]
\centering
\includegraphics[width=1\linewidth, trim=0mm 0mm 0mm 0mm]{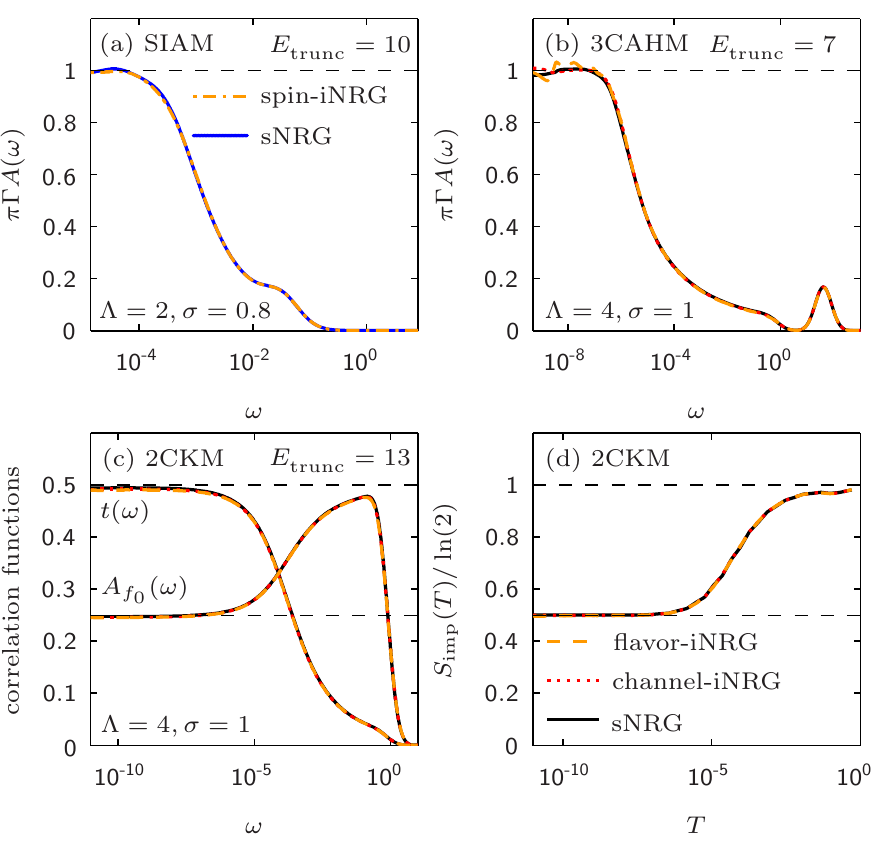} 
\caption{(Color online)
  Comparison of physical quantities calculated with iNRG and
  sNRG.  (a) Impurity spectral function $\pi\Gamma
  A(\omega)$ at $T=0$ for the SIAM; (b) impurity spectral
  function $\pi \Gamma A(\omega)$ at $T=0$ for the 3CAHM;
  (c) spectrum of the $t$ matrix $t(\omega)$ and the
  correlator $A_{\mathrm{f}_{0}}(\omega)$ at $T=0$ for the
  2CKM; (d) impurity contribution to the entropy,
  $S_{\mathrm{imp}}(T)$, for the 2CKM. For the dynamical
  correlators shown in (a)-(c), the protocol of
  Ref.~\onlinecite{Weichselbaum2007} was used to broaden
  discrete data, using a broadening parameter of
  $\sigma_{\rm broad} = 0.8$ or $1$ for $\Lambda = 2.0$ or
  4.0, respectively. All quantities were $z$-averaged over
  $z=0$ and $z=0.5$, except for the flavor-interleaved 3CAHM
  spectral function in (b), which was only calculated for
  $z=0$ ($z=0.5$ exceeded memory resources). The observed
  low-frequency oscillations are therefore an artifact of
  underbroadening, and would be removed by additional
  $z$ averaging or use of a larger $\sigma_{\rm broad}$.}
\label{fig_Paper_interleaved_2}
\end{figure}

The absolute accuracy of both iNRG and sNRG can be directly
assessed from calculated physical quantities. In particular,
we focus on $T=0$ correlation functions. For the
particle-hole symmetric SIAM and 3CAHM, the impurity
spectral function $A(\omega)=
-\tfrac{1}{\pi}\text{Im}~\langle{\langle \hat{d}_{\nu}} ;
{\hat{d}_{\nu}^{\dagger}}\rangle\rangle_{\omega}$ is pinned
by the Friedel sum rule at the Fermi level, $\omega=0$. The
exact analytic result\cite{hewson} is $\pi \Gamma A(0)=1$.
As a measure of the accuracy in NRG, we therefore consider
the relative deviation at the Fermi energy,
\begin{subequations}
\label{eq:deviations}
\begin{eqnarray}
   \delta A(0)/ A(0) = \pi \Gamma A(0)-1 , 
\end{eqnarray}
shown in Figs.~\ref{fig_Paper_interleaved_1}(j) and \ref{fig_Paper_interleaved_1}(l). For the
2CKM, we consider the spectrum $t(\omega)=-\pi \rho(\omega)
\text{Im}~T(\omega)$, where $T(\omega)$ is the scattering
$t$ matrix. Again, the spectrum is pinned at the low-energy
non-Fermi liquid fixed point; the exact analytic
result\cite{CFT2CK} is $t(0)=\tfrac{1}{2}$. In
Fig.~\ref{fig_Paper_interleaved_1}(k), we therefore consider
the relative NRG deviation at the Fermi energy,
\begin{eqnarray}
   \delta A(0)/ A(0) = 2t(0)-1 \;.
\label{eq:deviations_2ck}
\end{eqnarray}
\end{subequations}

We find that iNRG and sNRG perform similarly, recovering
exact results to within a few percent for $\Etrunc>7$. For
each case studied, iNRG appears to deviate somewhat stronger
from $\delta A(0)/A(0)= 0$ for $\Etrunc<7$ than sNRG; but
approximately equivalent results are obtained for
$\Etrunc>7$.  Even when interleaving all 6 flavors in the
3CAHM, using Abelian symmetries only, we similarly
anticipate that $\delta A(0)/A(0)$ will converge to $0$ for
sufficiently large $\Etrunc$. [This is supported in
Fig.~\ref{fig_Paper_interleaved_1}(l) by the orange data
point at $\Etrunc=7$, which was calculated for $z=0$ only.]

This conclusion is further substantiated by
Fig.~\ref{fig_Paper_interleaved_2}, which shows the full
frequency dependence of dynamical correlation functions at
$T=0$ in panels (a) - (c), and the temperature dependence of the
impurity entropy in panel (d). For the SIAM in panel (a),
the iNRG and sNRG impurity spectral functions are
essentially indistinguishable for $\Etrunc=10$ and
$\Lambda=2$, at all frequencies. For the 3CAHM in panel (b),
channel-iNRG and sNRG results for the impurity spectral
function are again indistinguishable for $\Etrunc=7$ and
$\Lambda=4$. Flavor-iNRG shows some oscillations on the
lowest energy scales due to underbroadening: the iNRG
calculation was performed only for $z=0$. Obtaining a
completely smooth curve would either require additional
$z$ averaging (but $z=0.5$ exceeded memory resources) or the
use of a larger broadening,  $\sigma_{\rm broad}$. Panel (c)
shows the spectrum of the $t$ matrix $t(\omega)$, and the
local bath spectral function $A_{f_0}(\omega) =
-\tfrac{1}{\pi}\text{Im}~\langle{\langle \hat{f}_{0\nu }} ;
{\hat{f}_{0\nu}^{\dagger}}\rangle\rangle_{\omega}$ for the
2CKM. At $\Etrunc=13$ for $\Lambda=4$, both iNRG and sNRG
yield equivalent and highly accurate results. Finally, panel
(d) confirms that thermodynamic quantities (here illustrated
for the impurity contribution to the total entropy) are
accurately reproduced using both iNRG and sNRG for the
2CKM. In particular, the non-trivial residual
entropy\cite{CFT2CK}
$S_{\text{imp}}(T=0)=\tfrac{1}{2}\ln(2)$ is correctly
reproduced.

In Figs.~\ref{fig_Paper_interleaved_1}(m)-\ref{fig_Paper_interleaved_1}(o), we examine the
convergence of both iNRG and sNRG calculations, analyzed
quantitatively in terms of the NRG discarded weight $\delta
\rho_{\text{disc}}$ (see \Sec{sec:discweight}).  As
expected, the discarded weight decays exponentially with
increasing $\Etrunc$. The calculations are considered fully
converged when $\delta \rho_{\text{disc}}<10^{-6}$, which is
reached in all cases at around $\Etrunc\approx 7$. No
qualitative changes occur in physical results on further
increasing $\Etrunc$ [panels (j)-(l)].
Figures~\ref{fig_Paper_interleaved_1}(m)-\ref{fig_Paper_interleaved_1}(o) show clearly that
the convergence behavior of iNRG is equivalent to that of
sNRG, implying that the states additionally discarded by
iNRG at intermediate steps do \emph{not} have appreciable
weight in the eigenstates of later iterations.  Indeed, the
discarded weights for iNRG (orange and red symbols) and sNRG
(green, blue and black symbols) for the same $\Lambda$ are
approximately equal.  [The only exception is seen in panel
(m), for $\Nchan= 1$ and $\Lambda = 4$, where the discarded
weight differences between sNRG (green circles) and iNRG
(orange circles) are apparently somewhat larger. We
attribute this to inaccuracies in the estimation of the
discarded weight, since, by far, the smallest number of data
points (diagonal weights $\rho_s$) were available for the
extrapolation in this case.]

For $\delta \rho_{\text{disc}} \gtrsim 10^{-6}$, i.e.,\ above
the convergence threshold,  the discarded weights behave
similarly for \textit{all} NRG calculations irrespective of
the choice of $\Lambda$; below this threshold, the behavior
becomes somewhat dependent on $\Lambda$: for a given
$\Etrunc$, larger $\Lambda$ yields a larger discarded weight
both for iNRG and sNRG [panels (m) and (n)]. The reason for this
is that the spectrum of \textit{rescaled} eigenenergies in
NRG shows a $\Lambda$ dependence for higher energies: while
rescaling is designed to ensure that the low-energy regime
(dominated by single-particle excitations) of the rescaled
eigenspectrum is almost $\Lambda$-independent, it stretches
apart the high-energy regime (dominated by many-particle
excitations). High-energy states are therefore shifted up
more for larger  $\Lambda$. The consequence is that, on
increasing $\Etrunc$ and $\Lambda$, the weight of the
reduced density matrices is shifted to higher rescaled
energies. This means that the slope, $\kappa$, of the dashed
red line in Fig.~\ref{fig_dw} would decrease, causing an
increase in the total integrated discarded weight, $\delta
\rho_{\text{disc}}$.


\begin{figure}[t]
\centering
\includegraphics[width=1\linewidth, trim=0mm 0mm 0mm 0mm]{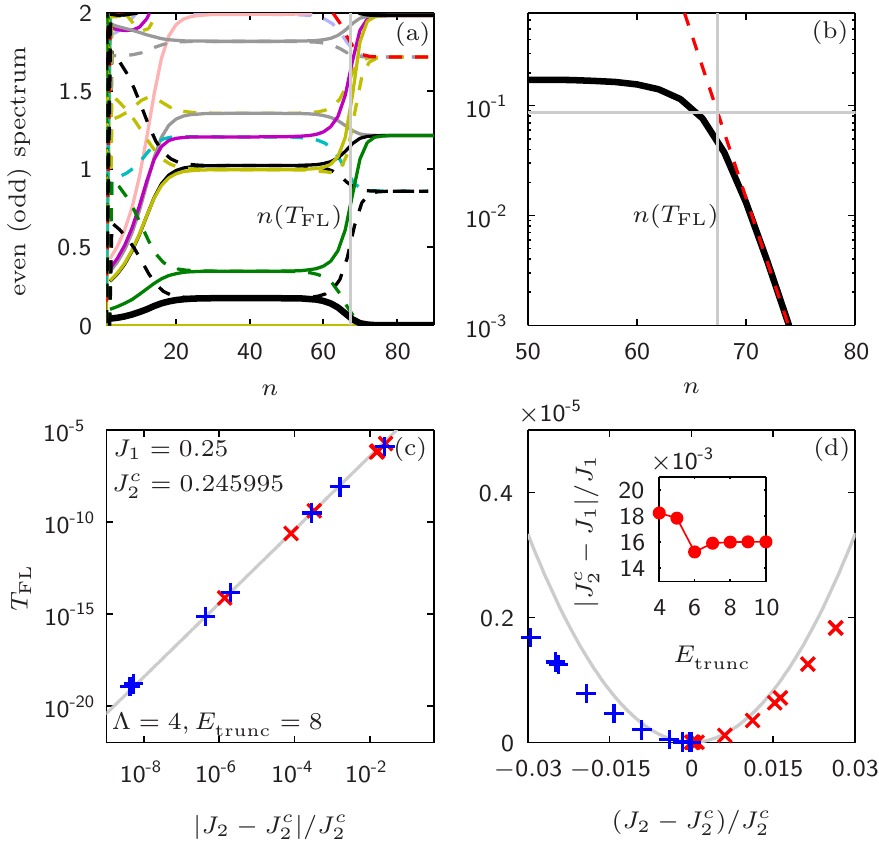} 
\caption{(Color online)
  Fine tuning in iNRG for the 2CKM (a) Flow of iNRG
  many-particle energies with Wilson shell index $n$ [solid
  (dashed) lines for even (odd) $n$] for the 2CKM. Different
  colors correspond to states with different quantum
  numbers.  (b) The Fermi liquid crossover scale,
  $T_{\text{FL}}$, can be extracted from the flow of the
  first excited state (thick black line): we fit its
  large-$n$ behavior with a power law (dashed red line),
  take $n(T_{\text{FL}})$ to be the iteration number
  [vertical grey line in (a) and (b)] at which this power
  law reaches half of the fixed-point value of this state
  (horizontal grey line), and define the Fermi liquid scale
  as $T_{\text{FL}}=\omega_{n(T_{\text{FL}})}$. In 
  (c) and (d), the resulting values of $T_{\rm FL}$ are
  plotted as function of $J_2 - J_2^c $ on a log-log or
  linear plot, respectively, using red (blue) symbols for
  $J_2 > J_2^c$ ($< J_2^c$). Grey lines give the asymptotic
  form $T_{\rm FL} \sim (J_2-J_2^c)^2$. By using an
  {extrapolative} protocol, the critical coupling $J_2^c$
  can be located exponentially rapidly in the number of
  separate iNRG runs. [Inset to (d)] The difference between
  the critical coupling $J_2^c$ and $J_1$, plotted as a
  function of the truncation energy.
}
\label{fig_Paper_interleaved_3}
\end{figure}


\subsection{Fine tuning in iNRG}\label{sec_tuning}

If a given model possesses an exact flavor symmetry -- and
furthermore, if the breaking of this flavor symmetry is an
RG \emph{relevant} perturbation -- iNRG must be combined
with parameter fine tuning. This is because the asymmetric
discretization required to interleave different Wilson
chains in iNRG artificially breaks bare flavor symmetries,
albeit rather weakly.  However, effective channel symmetry
in the discretized model can be restored through the
fine tuning of couplings.\cite{Mitchell2014}

A prime example is the 2CKM, for which channel
symmetry-breaking is relevant.\cite{Nozieres,Affleck1992}
The critical point of the 2CKM is realized at precisely
$\rho_{1}(0)J_1 = \rho_{2}(0)J_2$, embodying the frustration
responsible for its non-Fermi liquid properties. In sNRG,
channel symmetry is exactly preserved:
$\rho_{1}^{\text{disc}}(\varepsilon) =
\rho_{2}^{\text{disc}}(\varepsilon)$, and so the critical
physics is accessible along the line $J_1=J_2$ (only the
Kondo temperature $T_{\text{K}}^{\text{2CK}}$ is affected by
the actual value chosen for $J_1=J_2$).  However, we note
that even in sNRG, the precise value of
$\rho^{\text{disc}}_{\alpha}(0)J_{\alpha}$ can deviate very
slightly from the bare value $\rho_{\alpha}(0)J_{\alpha}$,
due to the discretization.  Although the
$T_{\text{K}}^{\text{2CK}}$ obtained in sNRG might therefore
also be slightly different from the true value, it should be
emphasized that the \emph{universal} low-energy physics is
identical.

Likewise, $\rho^{\text{disc}}_{\alpha}(0)J_{\alpha}$
deviates from $\rho_{\alpha}(0)J_{\alpha}$ in iNRG. However,
the important difference is that
$\rho^{\text{disc}}_{1}(0)J_{1}\ne
\rho^{\text{disc}}_{2}(0)J_{2}$, even when
$\rho_{1}(0)J_{1}= \rho_{2}(0)J_{2}$. In the presence of
this small channel asymmetry perturbation, the critical
point is destabilized, leading to a flow \emph{away} from
the non-Fermi liquid fixed point, and toward a stable Fermi
liquid fixed
point.\cite{Affleck1992,CFT2CK,akm:exactNFL,*akm:finiteT}
The temperature/energy scale characterizing this Fermi
liquid crossover is denoted $T_{\text{FL}}$.  To access the
critical physics for a given $J_1$, one must therefore
fine tune the value of $J_2\rightarrow J_2^c$ such that
$T_{\text{FL}}\rightarrow 0$.  In principle, $T_{\text{FL}}$
can be extracted from any physical quantity; it can also be
extracted directly from the flow of NRG many-particle
energies, as shown in
Figs.~\ref{fig_Paper_interleaved_3}(a) and \ref{fig_Paper_interleaved_3}(b) (see caption for
details).

A very efficient extrapolative tuning protocol can be
employed if the functional dependence of $T_{\text{FL}}$ on
$J_2-J_2^c$ is known analytically. In the case of the 2CKM,
it is known\cite{CFT2CK} that $T_{\text{FL}}\sim
(J_2-J_2^c)^2$ when $T_{\text{FL}}\ll
T_{\text{K}}^{\text{2CK}}$.  This can be exploited by
adopting the following protocol (somewhat similar to
Newton's method for finding roots from a linear fit): the
lowest two values of $T_{\text{FL}}$ extracted from previous
iNRG runs are used to fit a parabola; the trial value of
$J_2$ for the next iNRG run is then given by the minimum of
the parabola.  This protocol is illustrated in
Figs.~\ref{fig_Paper_interleaved_3}(c) and \ref{fig_Paper_interleaved_3}(d). $J_2$ converges to
the critical value $J_2^c$   exponentially rapidly in the
number of separate iNRG runs. In
Fig.~\ref{fig_Paper_interleaved_3}(c), $T_{\text{FL}}$
decreases by roughly one order of magnitude per iNRG run.

When the dependence of $T_{\text{FL}}$ on the model
parameters is not known analytically, a more general
bisection method can instead be used to locate the critical
point, provided the two phases separated by it can be
distinguished in different iNRG runs. For example, in the
2CKM, the critical point $J_2^c$ separates Kondo strong
coupling phases where the impurity spin is ultimately fully
screened by either lead $\alpha=1$ or $2$ (depending on the
sign of $J_2-J_2^c$). These phases can be distinguished by
physical observables, e.g.,\ the $t$ matrix for channel
$\alpha$, since $t_{\alpha=2}(0)=1$ and $t_{\alpha=1}(0)=0$
when $J_2>J_2^c$.  In practice, a simpler and more direct
way to distinguish the two phases involves comparing their
NRG fixed point energy level structures, which are indexed
differently.

The bisection method also involves multiple iNRG runs: each
new run uses a value $J_2$ that is an average of two
previous $J_2$ values (one in each phase) lying closest to
each other.  $T_{\text{FL}}$ does not need to be calculated
explicitly here.  This protocol also locates the critical
point exponentially rapidly (although utilizing information
about the functional dependence of $T_{\text{FL}}$, where
available, is the optimal strategy).

Finally, we note that the precise value of $J_2^c$ in iNRG
depends on the discretization details. However, the critical
ratio $J^c_2/J_1$ is generally found to deviate from its
exact (undiscretized) value of $1$ by about 1\% [see the
inset of Fig.~\ref{fig_Paper_interleaved_3}(d)].  We also
find that $J^c_2/J_1$ converges to a specific value on
increasing $\Etrunc$, and is essentially invariant for
$\Etrunc> 7$. This indicates that the critical value of
$J^c_2$ determined by the above tuning protocol in iNRG is
the true (converged) value \emph{for the asymmetrically
discretized model}.


\section{Conclusion}

In this paper, we compared two methods for treating
multiband quantum impurity problems with NRG: sNRG
exploiting model symmetries,\cite{Weichselbaum2012} and iNRG
exploiting symmetry-breaking.\cite{Mitchell2014} 

Our analysis of the NRG discarded
weight\cite{Weichselbaum2011} and the error in certain
calculated physical quantities demonstrates that sNRG and
iNRG are of comparable accuracy when the same discretization
parameter $\Lambda$ is used, and when the same number of
states are kept on average at each iteration. iNRG therefore
constitutes a more fine-grained RG scheme, in which
intermediate state-space truncations do not adversely affect
convergence or accuracy.

For models that possess high intrinsic symmetries, sNRG is a
highly efficient tool for treating multiband quantum
impurity problems, because full use can be made of the
symmetries. But in models with lower symmetries, sNRG
quickly becomes inefficient, and in practice unusable, when
more than two spinful conduction electron channels are
involved.

We find that iNRG is much more efficient than sNRG for
treating a given model with equivalent settings. This is the
appropriate comparison for systems where bare model flavor
symmetries are already broken. Such a scenario naturally
arises on inclusion of a magnetic field, potential
scattering, channel anisotropies, and in the \emph{vicinity}
of high-symmetry critical points. In these cases, iNRG has
the clear advantage.

For high-symmetry models where sNRG can exploit larger
symmetries than iNRG, the `best' sNRG and iNRG calculations
are found to be of roughly comparable efficiency. In this
case, iNRG can be regarded as a viable and technically
simple alternative to sNRG.

However, \emph{optimal} efficiency can often be obtained by
\emph{combining} features of sNRG and iNRG to interleave the
Wilson chains for some electronic flavors, while retaining
and exploiting other symmetries. 

The results of this paper suggest that iNRG could find
powerful application as an impurity solver for multiband
DMFT. For example, Hubbard models of transition metal oxides
with partially filled $d$ orbitals, ruthenates, or iron
pnictide and chalcogenide high-temperature superconductors
map within DMFT to effective multi-channel impurity problems
that could be solved accurately using iNRG. In the context
of simulating real strongly correlated materials, channel
symmetries are generally broken (for example, due to crystal
field splitting).  Our analysis indicates the feasibility of
studying such channel-asymmetric models for three effective
channels, and further suggests that 4- and even 5-channel
problems could be tackled using iNRG in the future.

We conclude that iNRG is a competitive and versatile
alternative to sNRG, even for high-symmetry models. When
large symmetries are not available, iNRG is far more
efficient that sNRG. Moreover, iNRG provides a way forward
for complex models with lower symmetries that are beyond the
reach of sNRG, opening up possibilities for new
applications of NRG as an impurity solver.


\acknowledgements 
KMS, AW, and JvD were supported by the DFG through
SFB-TR12, SFB631, WE4819/1-1, WE4819/2-1, and the Cluster of
Excellence \textit{NanoSystems Initiative Munich}.
AKM acknowledges funding from the D-ITP consortium,
a program of the Netherlands Organisation for Scientific
Research (NWO) that is funded by the Dutch Ministry
of Education, Culture and Science (OCW).


\begin{appendix}
\section{Choice of truncation energy in iNRG} \label{appendix-truncation}

In this appendix, we provide a heuristic justification of the choice of truncation energy $E_{\rm trunc}^{\iNRG}$ proposed in Eq.~(\ref{eq:Ectrunc}).

\begin{figure}[b]
\centering
\includegraphics[width=1\linewidth, trim=0mm 0mm 0mm 0mm]{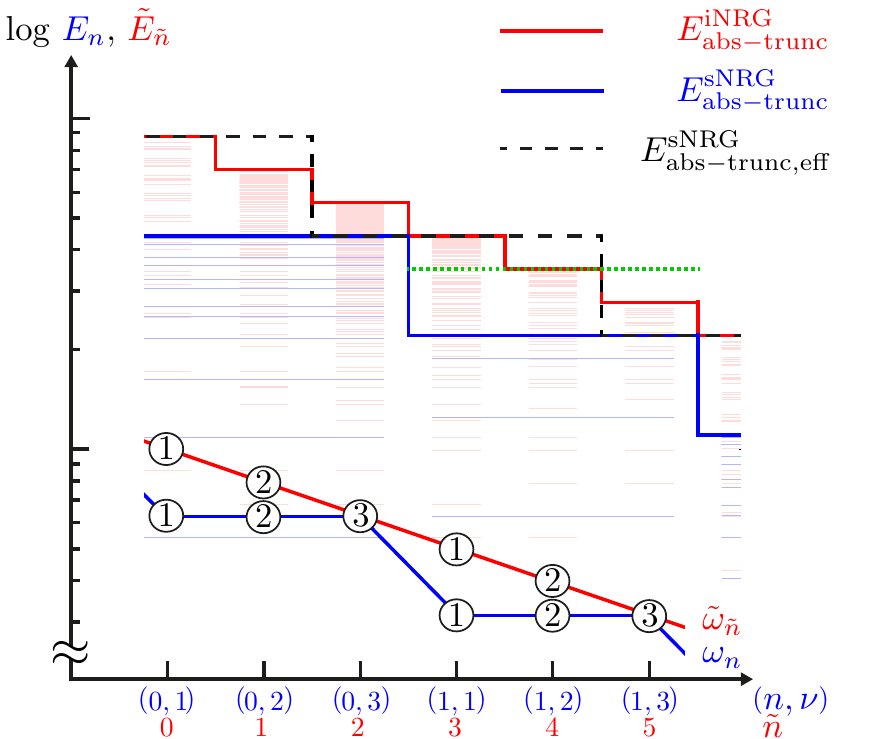} 
\caption{(Color online)
  Schematic depiction of the sNRG (blue) and iNRG (red)
  truncation schemes used here, illustrated for a model with
  $m=3$ flavors.  The vertical axis corresponds to absolute
  energies on a logarithmic scale, while the Wilson (sub-)shell
  index is given on the horizontal axis.  The lower part of
  the sketch depicts the evolution of the characteristic
  energies $\omega_n$ and $\tilde \omega_{\tilde n}$ for
  sNRG and iNRG.  Thin faint blue and red lines depict the
  excitation eigenenergies (relative to the ground state
  energy) of sNRG supershells or iNRG subshells. In sNRG,
  all three subsites comprising the supersite for that
  iteration are added at once (there is no intermediate
  truncation), while in iNRG the subsites are added
  separately, and truncation occurs at each step.  The
  absolute truncation energies
  $\text{E}_{\text{abs-trunc}}^{\sNRG}$ and
  $\text{E}_{\text{abs-trunc}}^{\iNRG}$ therefore form two
  different staircases, depicted as the thick blue and red
  lines, respectively (the step width for sNRG is $m$ times
  longer than that of iNRG). States with higher energies are
  discarded.  The truncation pattern of sNRG, when viewed
  from the perspective of \iNRG, amounts to employing the
  effective truncation energy
  $\text{E}_{\text{abs-trunc,eff}}^{\sNRG}$, shown as the
  black dashed line: by using a high truncation threshold
  (that of the previous iteration,
  $\Etrunc^{\sNRG}\times\omega_{n-1}$) for the first $m-1$
  subsites, and then dropping to
  $\Etrunc^{\sNRG}\times\omega_{n}$ only for the last
  subsite, all states are effectively kept until the
  supersite is complete. Viewed from this iNRG perspective,
  the truncation energies of iNRG and sNRG are the same on
  average (green dotted  line for supersite $n=1$) and the
  areas under the red solid, black dashed, and green dotted
  lines are the same, provided $\Etrunc^{\sNRG}$ and
  $\Etrunc^{\iNRG}$ are related via Eq.~(\ref{eq:Ectrunc}).
}
\label{fig_Etrunc}
\end{figure} 

In iNRG, the subsites $\tilde n = (n, \nu)$ of supersite $n$
are added one by one, each followed by a truncation with a
different absolute truncation energy,
$\text{E}_{\text{abs-trunc}}^{\iNRG}=\Etrunc^\iNRG \tilde
\w_{(n,\nu)}$.  The geometric average of these truncation
energies over the supershell is
\begin{align}
  \label{sec:Etrunc-abs-geom-iNRG}
  \langle \Eabstrunc^\iNRG \rangle_n^\geom = 
  \Etrunc^\iNRG \left({\textstyle \prod_{\nu=1}^\Nflavor }
\, \tilde \w_{(n,\nu)} \right)^{1/\Nflavor} \, . 
\end{align}

In sNRG, by contrast, all $\Nflavor$ subsites of supersite
$n$ are added as one unit, followed by truncation at the
absolute truncation energy
$\text{E}_{\text{abs-trunc,eff}}^{\sNRG}=\Etrunc^\sNRG \w_n $.
The thick red and blue
lines in Fig.~\ref{fig_Etrunc} show the resulting evolution
of the absolute truncation energies in iNRG and sNRG with NRG iteration
number, respectively. The characteristic energies $\tilde{\w}_{\tilde{n}}$ and
$\w_n$ are shown as the circles in the
lower part of the figure.

To meaningfully compare sNRG and iNRG, it is
instructive to view the truncation profile of sNRG within
the framework of iNRG. One can think of sNRG as an effective
iNRG calculation, in which subsites are added separately,
but the effective truncation threshold
$\text{E}_{\text{abs-trunc,eff}}^{\sNRG}$ for the first
$m-1$ subsites is high enough so that all states are kept.
This is guaranteed by using the absolute truncation energy
of the previous iteration,
$\text{E}^{\sNRG}_{\text{trunc}}\times \omega_{n-1}$. Only
when the supersite is complete after adding the last subsite
with $\nu=m$, the effective absolute  trucation energy is
reduced to induce the necessary truncation
$\text{E}^{\sNRG}_{\text{trunc}}\times \omega_{n}$. Overall,
the effective truncation energy in sNRG is
subsite-dependent: specifically, within supersite $n$, we
have $\text{E}_{\text{abs-trunc,eff}}^{\sNRG}=\Etrunc^\sNRG
\ttimes \w_{n-1+\delta_{\nu m}}$. This is shown as the black
dashed line in Fig.~\ref{fig_Etrunc}.  The geometric average
of the effective sNRG truncation energies is
\begin{align}
\label{sec:Etrunc-abs-geom-sNRG}
  \langle \Eabstrunc^\sNRG \rangle_n^\geom = 
  \Etrunc^\sNRG \left(\w_n {\textstyle \prod_{\nu=1}^{m-1}}
  \w_{n-1} \right)^{1/m} \, . 
\end{align}
By demanding that the average truncation energies
Eqs.~(\ref{sec:Etrunc-abs-geom-iNRG}) and
(\ref{sec:Etrunc-abs-geom-sNRG}) are the same (illustrated
by the green dotted line in Fig.~\ref{fig_Etrunc} for
iteration $n=1$), we obtain the relation between
$E_{\text{trunc}}^{\iNRG}$ and $E_{\text{trunc}}^{\sNRG}$
announced in Eq.~(\ref{eq:Ectrunc}). 

Finally, we comment that, given a specific number of flavors
$m$, the choice of Eq.~(\ref{eq:Ectrunc}) implies that the
area under the lines $\text{E}_{\text{abs-trunc}}^{\iNRG}$
(red) and $\text{E}_{\text{abs-trunc,eff}}^{\sNRG}$ (black
dashed) is the same for each supersite $n$ and (as
examplified for $n=1$ in Fig.~\ref{fig_Etrunc}) corresponds
to the area under the green dotted line.

The important consequence of effectively using `same' absolute
truncation energies on average is that the number of kept states turns
out to be the same on average for iNRG and sNRG. Nevertheless, similar
to even-odd effects in the number of states of sNRG,
subshell-dependent variations of $\Nkept$ occur in iNRG (see
Fig.~\ref{fig_Paper_interleaved_5}).  


\section{Discarded weight based on energy eigenstates}
\label{appendix-discarded}

In this appendix, we describe how to quantify the contributions of highlying \textit{energy eigenstates} to reduced density matrices, rather than evaluating the eigenspectrum of the reduced density matrices as in Ref.~\onlinecite{Weichselbaum2011}.
    
\begin{figure}[tb]
\centering
\includegraphics[width=1\linewidth, trim=0mm 30mm 0mm 0mm]{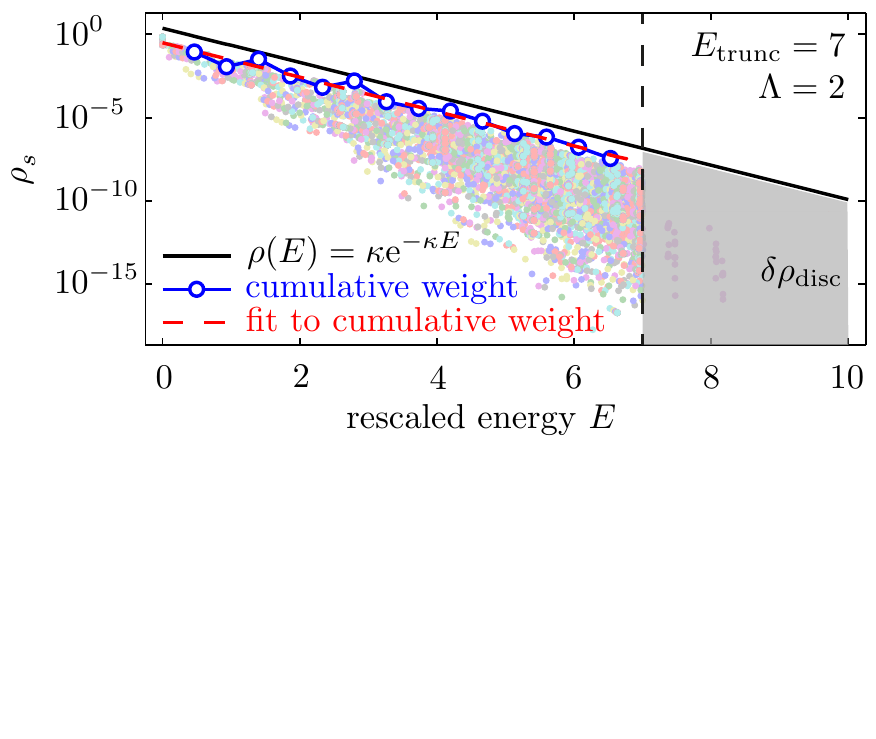} 
\caption{(Color online)
  Estimating the discarded weight, $\delta\rho_{\rm disc}$,
  of a single NRG run. Colored dots show the diagonal
  weights of the reduced density matrices in the energy
  eigenbasis of NRG. Different colors represent the weights
  for different NRG iterations. We calculate and plot the
  cumulative weights (blue open circles) using $16$ bins in
  the energy window $[0, \Etrunc]$. The truncation energy
  $\Etrunc=7$ is indicated by the vertical black dashed
  line. The red dashed line is an exponential fit to the
  cumulative weights; its slope gives $\kappa$ as defined
  in Eq.~(\ref{eq:rhodisc}). The  black line shows  the
  normalized integrated weight distribution $\rho(E)=\kappa
  {\rm e}^{-\kappa E}$, extrapolated to energies
  $E>\Etrunc$. The shaded grey area under this black line
  then serves as estimate for the discarded weight:
  $\delta\rho_{\rm disc}={\rm e}^{-\kappa \Etrunc}$. This
  example is well-converged, with $\kappa=2.37$ yielding
  $\delta\rho_{\rm disc}=6.23\times 10^{-8}$.
}
\label{fig_dw}
\end{figure}

We do this by analyzing the diagonal weights of $\hat\rho$,
i.e.,\ the diagonal elements of the reduced density matrix in
the \emph{energy} eigenbasis $|s\rangle_{\rm{K}}$ within the
kept sector, $\rho_s=\vphantom{\rangle}_{\rm{K}}\langle s|
\hat\rho|s\rangle_{\rm{K}}$.  Hence we employ a
strategy analogous to that of
Ref.~\onlinecite{Weichselbaum2011}, but here we 
use the energy eigenbasis (cf.\ Fig.~3 of
Ref.~\onlinecite{Weichselbaum2011} and Fig.~\ref{fig_dw})
rather than the eigenbasis of the reduced density matrices
(cf.\ Fig.~4 of Ref.~\onlinecite{Weichselbaum2011}) to
estimate the discarded weight. This leads to a slightly
different definition of the discarded weight, as described
below.

Due to the energy scale separation in NRG, the diagonal weights of the
reduced density matrices decrease exponentially when plotted versus
their corresponding rescaled eigenenergies (cf.\ colored dots in
Fig.~\ref{fig_dw}).  The same is true for the integrated weight
distribution [cf.\ Eq.~(18) of Ref.~\onlinecite{Weichselbaum2011} and
black solid line in Fig.~\ref{fig_dw}], which as such constitutes an
upper bound for the weights, and scales as
$\rho(E)\approx\kappa {\rm e}^{-\kappa E}$ (normalized such that
$\int_0^\infty \rho(E)\,dE =1$). This exponential decay shows that the
contribution of an NRG state with rescaled energy $E$ to the
properties of subsequent shells decreases exponentially with $E$. This
justifies the strategy in NRG to keep track of these contributions
only up to a threshold energy of $\Etrunc$. Moreover, by extrapolating
the exponential form to energies beyond $\Etrunc$, the sum of weights
associated with all discarded high energy states with $E > \Etrunc$
can be estimated.  We therefore define the total discarded weight by
the following integral (represented by the shaded grey area in
Fig.~\ref{fig_dw}):
\begin{equation}
\label{eq:rhodisc}
  \delta\rho_{\rm disc}=\int_{\Etrunc}^{\infty}\rho(E) \, dE 
  ={\rm e}^{-\kappa \Etrunc} \;.
\end{equation}%

In practice, we obtain $\delta\rho_{\rm disc}$ numerically
as follows.  First, a cumulative histogram is constructed of
the discrete weights $\rho_s$ for $E<\Etrunc$ over all NRG
iterations, using coarse-grained energy bins (e.g.,\ keeping
$16$ bins in the energy window $[0, \Etrunc]$).  This
histogram represents $\rho(E)$. A linear fit to its shape on
a semi-logarithmic scale then yields $\kappa$, which in turn
gives $\delta\rho_{\rm disc}$, via \Eq{eq:rhodisc}. Since
$\delta\rho_{\rm disc}$ depends only on the dimensionless
quantity $\kappa \Etrunc$, the result is independent of the
choice of energy unit for $\Etrunc$. 

By examining the decay of the discarded weight $\delta
\rho_{\rm disc}$ with increasing $\Etrunc$, and observing
the corresponding convergence of physical quantities, we
have found that calculations can be considered converged
when $\delta\rho_{\rm disc}<10^{-6}$. This convergence
criterion has been determined to ensure that further
increasing $\Etrunc$ does not qualitatively change the
results [see, e.g.,\ Figs.~\ref{fig_Paper_interleaved_1}(j)-\ref{fig_Paper_interleaved_1}(o)].  We note that the numerical value of this threshold is 
about the square-root of and thus
considerably larger than that reported in
Ref.~\onlinecite{Weichselbaum2011}, which was obtained using
the alternative definition of discarded weight in terms of
the eigenspectrum of reduced density matrices.

An important advantage of defining the discarded weight in
terms of the energy eigenbasis is that $\delta \rho_{\rm
disc}$, evaluated at fixed $\Etrunc$, is rather insensitive
to changing the discretization parameter $\Lambda$ (we
verified this explicitly over a range of $\Lambda$ typically
used in NRG, $1.7 \lesssim \Lambda \lesssim 7$). We found
that $\Etrunc \gtrsim 7$ generally suffices to obtain
well-converged results for physical quantities.  In
contrast, the discarded weight defined in terms of density
matrix eigenvalues\cite{Weichselbaum2011} turns out to show
a much more pronounced dependence on $\Lambda$ at fixed
$\Etrunc$, which would be inconvenient for the present
purposes.

\end{appendix}
  


%


\end{document}